\documentclass[seceq]{ptptex}

\usepackage{graphicx}
\usepackage{wrapft}
\usepackage{fancybox}
\newcommand{\la}{\langle}
\newcommand{\ra}{\rangle}

\newcommand{\beq}{\begin{eqnarray}}
\newcommand{\eeq}{\end{eqnarray}}
\renewcommand{\d}{\partial}
\newcommand{\bfl}{\begin{flushleft}}
\newcommand{\efl}{\end{flushleft}}
\notypesetlogo                       %comment in if to eliminate PTPTeX 
\title{%        %You can use \\ for explicit line-break
Physical Origin of Chiral States and \\
Near-Threshold 
Resonances Observed at BES
}

\author{%       %Use \scshape  for the family name
Shin \textsc{Ishida}
}

\inst{%         %Affiliation, neglected when [addenda] or [errata]
Research Institute of Science and Technology, 
College of Science and Technology, \\
Nihon University, Tokyo 101-8308, Japan
}

\abst{%         %this abstract is neglected when [addenda] or [errata]
The purpose of my talk is to explain physical backgrounds of 
the chiral states, which are predicted to exist generally in the 
low mass regions of hadrons, containing light quarks, and to 
stimulate the experimental and phenomenological search for their candidates.
The contents are organized as follows: \\

\S 1. Present Status of Hadron Spectroscopy\\

\S 2. Some Basic Consideration on Quark Confinement\\

\S 3. Essentials of Description of Composite Hadron in $\widetilde{U}(12)$-Scheme\\

\S 4. Revisal of Meson and Baryon Wave Functions and Chiral States\\

\S 5. Multi-Quark Hadrons and Near-Threshold Resonances\\

\S 6. Concluding Remarks

}
\begin{document}
\maketitle
%%%%%%%%%%%%%%%%%%%%%%%%%%%%%%%%%%%%%%%%%%%%%%%%%%%%%%%%%%%%%%%%%%%%%%%%%
%\begin{flushleft}
%%%% NO2  %%%%%%%%%%%%%%%%%%  

\section{Present Status of Hadron Spectroscopy}

Now hadron spectroscopy seems to meet a revolutionary stage. There exist 
\vspace{0.2cm}

\let\tabularsize\normalsize
%%%%%%%%%%%%%%%%%%%%
\bfl
({\it Two Contrasting Views for Classification of Hadrons}),
\efl
\begin{table}[h]
\begin{center}
\caption{Successes and Difficulties of the View-points}
\begin{tabular}{l c|c} 
\hline
\hline
& Non-Relativistic View & Relativistic View \\
\hline
Bases & Non-Relativistic Quark Model & Field-Theoretical Models \\ 
Framework & $SU(6)_{SF}\otimes O(3)_{L}$ & Spont. Broken Chiral Symm.\\ 
Difficulty & Lacking Lorentz Covariance, ~~~~~~~ & Unable to treat~~~~~~~\\ 
&Out of Notion of Chiral Symmetry &~Internal Excitations\\
\hline
\end{tabular}
\label{tab1}
\end{center}
\end{table}
\begin{flushleft}
which, on the one hand, have respective successes and, on the other hand, 
include their own difficulties as shown in Table \ref{tab1}.
\end{flushleft}

\vspace{0.3cm}
\bfl
({\it Observation of New ``Exotic'' Hadrons})\\
\efl
~~~~Meanwhile, new hadrons, seemingly out of conventional classification scheme, 
have been reported{\cite{Swanson}} to exist successively with the 
%
%\vspace{0.1cm}
\bfl
{\it Features} of new exotic hadrons; 
\efl
\begin{description}
%\vspace{-0.2cm}
\item[(F1)] \ \ 
Mass, close to thresholds of their observed or intermediate decay channels.

\item[(F2)] \ \ 
Decay width, unexpectedly narrow.

\end{description}
Several years ago, corresponding to this situation, we made a 
\vspace{0.2cm}
%%%  NO3   %%%%%%%%%%%%%%
\bfl
({\it Proposal of Covariant $\widetilde{U}(12)_{SF}$-Classification Scheme})
{\cite{U12, Ds}}
\efl
which is an unification, keeping their succesful points and 
getting rid of the difficulties of the above two views. 

\vspace{0.3cm}
\bfl
{\it Framework }:
\efl
\begin{flushleft}
~~~~It has the following symmetry in the rest frame of hadrons, which is 
embedded in the Lorentz-covariant space as, 
\end{flushleft}
\begin{eqnarray}
\mbox{Hadron}-\left[ \begin{array}{cl}
\mbox{at rest}(\mbox{\bf{P}}=0), & \mbox{Static $U(12)_{SF} \otimes O(3)_{L}$-Symm.} 
\ \ \ {\rm  }\\
&\\
\mbox{moving}(\mbox{\bf{P}} \not= 0), & \mbox{Covariant $\widetilde{U} (12)_{SF} 
\otimes O(3,1)_{{\rm \scriptscriptstyle Lorentz} }$-space \ \ \ ;}\\
\end{array} \right.
\end{eqnarray}
where the above groups contain their sub-groups as follows: 
\beq
\widetilde{U}(12)_{SF} &\supset& SU(3)_{F} \otimes 
\widetilde{U}(4)_{\rm Dirac \ Spinor},\nonumber\\
\widetilde{U}(4)_{\rm D.S.}\stackrel{{\bf P}=0}{\longrightarrow}U(4)_{S}&\supset& 
SU(2)_{\sigma} \otimes SU(2)_{\rho} \ \ 
({\rm Dirac} \ \gamma \equiv \sigma \times \rho).
\eeq

The remarkble point in this scheme is that it contains a new approximate symmetry
{\cite{Ds}}$SU(2)_{\rho}$ for ``Confined-Light Quarks'', 
and the non-relativistic symmetry
 $SU(6)_{SF} \left[ \supset SU(3)_{F}\otimes SU(2)_{\sigma} \right]
\ {\rm is \ extended \ into} \ SU(12)_{SF}(\supset SU(3)_{F}\otimes SU(2)_{\sigma}
\otimes SU(2)_{\rho})$.

\vspace{1em}
The basic vectors for the new $SU(2)_{\rho}$ space consist of 
$\{ \Phi_{+,\alpha}(X), \Phi_{-,\alpha}(X) \}$
\footnote{ 
Here we concern only the transformation properties of hadron Wave Function(WF) 
related with the C.M. coordinates and simply write as $\Phi(X, r, \cdots)\to 
\Phi(X)$, where $X/r$ is center of mass /relative coordinate of hadrons.
}, which have, respectively the quantum numbers and are called
{\footnote{
The notion of urciton is introduced in Ref.~\citen{Ur}. It is a dimensionless
 Dirac spinor simulating confined quarks inside hadrons.}}
, as  
\beq
\left[ \begin{array}{ccc}
\Phi_{+,\alpha}(X): & \mbox{\rm Pauli Ur-(ex)citon}
&
j^{P}={\frac{1}{2}}^{+} \ \rho_{3}=+1\\
\Phi_{-,\alpha}(X):& \mbox{\rm Chiral Ur-(ex)citon}&
j^{P}={\frac{1}{2}}^{-} \ \rho_{3}=-1 \ .\\
\end{array} \right.
\eeq
They are connected mutually by{\cite{Watanabe}} 
\beq
\mbox{\rm Chirality Transformation}; \ \ 
\Phi^{\chi}_{\pm}(X)=\Phi_{\mp,\alpha}(X)=
\left[ -\gamma_{5} \Phi_{\pm}(X) \right].
\eeq
The chiral urciton $\Phi_{-,\alpha}$ with exotic quantum number 
($j^{P}={\frac{1}{2}}^{-}$) leads to Chiral States, a new-type of 
``Exotic Hadrons''. 

\vspace{2em}

%%% NO4  %%%%%%%%%%%%%%%%%%%%%%%%%%%%%%%%%%%%%%%%%%%%%%%%
\section{Some Basic Considerations on Quark-Confiment}
%%%%%%%%%%%%%%%%%%%%%%  

First we recapitulate the definition of and related formulas to 

\vspace{1em}
\bfl
({\it Lorentz Covariance for Local Spinning Particle}).  
\efl
~~~~For Lorentz transformation of space-time coordinates of 
``local spinning hadron''
\beq
X_{\mu} \to X_{\mu}^{'}=\Lambda_{\mu\nu}X_{\nu}=
(\delta_{\mu\nu}+\epsilon_{\mu\nu})X_{\nu}
\eeq
the hadron wave function generally of multi-component is 
transformed as
\beq
\Phi(X)\to \Phi^{'}(X^{'})=S(\Lambda)\Phi(X), \ \ 
S(\Lambda)\equiv (1+\frac{i}{2} \epsilon_{\mu\nu}\Sigma_{\mu\nu}), 
\eeq
where the $S(\Lambda)$ are relevant matrices to its spin and 
the generators $\Sigma_{\mu\nu}$ are 
classified into those for two sub-groups as 
\beq
&&{\rm for \ Rotation}\ \ \ \ \ \ \ 
J_{i}\equiv\frac{1}{2}\epsilon_{ijk}\Sigma_{jk}~, \ \ \ 
{\rm for \ Boost} \ \ \ \ \ \ \  K_{i}\equiv i \Sigma_{i4}~~. \nonumber\\
\ \ \ &&\mbox{({\rm unitary \ group})} 
~~~~~~~~~~~~~~~~~~~~~~~~
\mbox{(\rm non-unitary group)} \nonumber
\eeq
{\it Spin-1/2 case} \ \ \  The relevant concrete formulas 
in the case of Dirac field %with spin $1/2$ $\Phi_{\alpha}(X)$ 
are given as 
\begin{table}[h]
\begin{center}
\begin{tabular}{cll}
{\large 
\ \ \ }&{\large $J_{i}=\frac{1}{2}\sigma_{i}\otimes\rho_{0}$, \ \ \ \ } & 
\large{ 
$K_{i}= \frac{i}{2}\rho_{1}\otimes \sigma_{i}$ ,}\\
{\large \ \ \ }& 
{\large 
$S(\Lambda)=R({\bf \omega})={\rm e}^{-i{\bf \omega}{\bf J}}\ \ \ \ \ \ \ \ \ $}
&{\large  $S(\Lambda) = B(\chi)={\rm e}^{-i{\bf b}\cdot {\bf K}}$, 
}\\
{\large \ \ \ }&& 
{\large  $({\bf b}=\chi \hat{\bf v} ; ~~\tanh \chi =|{\bf v}| , 
{\bf v} \equiv{{\bf P}}/P_{0}$) }\\
\end{tabular}
\end{center}
\end{table}
\\
where $P_{\mu}$ is the four-momentum of Dirac particle. 
%It is to be nated that only $\sigma$-WF is described by 
%Pauli Spinor (with two components), while both $\sigma$- and $\rho$-WF is 
%described by covariant Dirac spinor (with four components). 
Here it should be stressd that the $SU(2)_{\rho}$-freedom
\footnote{
$ \Phi_{\alpha}=(\phi_{1} \ \phi_{2} \ \chi_{1} \ \chi_{2})^{t}
$ ,The relevant Dirac $\gamma$-matrices are decomposed into 
$\rho$ and $\sigma$ matrices as 
$\gamma({\rm 4 \ by \ 4}) \equiv \rho ({\rm 2 \ by \ 2}) 
\otimes \sigma ({\rm 2 \ by \ 2})$, 
where the $\sigma$ operating on $SU(2)$-space expanded by 
$(\phi_{1} \ \phi_{2})$ on $(\chi_{1} \ \chi_{2})$ and the 
$\rho$ operating on $SU(2)$-space expanded by $(\phi \ \chi)$, respectively. 
}
 is indispensable for Lorentz covariance. 
\vspace{0.3cm}\\
{\it Higher spin case} \ \ \  The WF of particle 
with higher-spin is given as the tensor with multi-Dirac indices 
$\Phi_{\alpha_{1}\alpha_{2}\cdots\alpha_{N}}(X)$, of which generator 
is given as a sum of relevant ones for respective indices. 
\vspace{0.6cm}
%%%  NO5  %%%%%%%%%%%%%%%
\bfl
({\it Hole Theory for Negative Energy solution of Dirac Equation })
\efl
~~~~As is well known, the Lorentz covariant Dirac equation with 
spin one-half leads necessarily to 
the negative-energy solutions, which are generally interpreted, 
applying the Hole Theory, as representing the anti-particle of 
relevant particles. However, contrary to the free electrons, 
application of H.T. to the confined quarks inside hadrons produces the 
difficulty as follows: \\
\begin{figure}[h]
\begin{center}
\includegraphics[width=12cm,height=10cm]{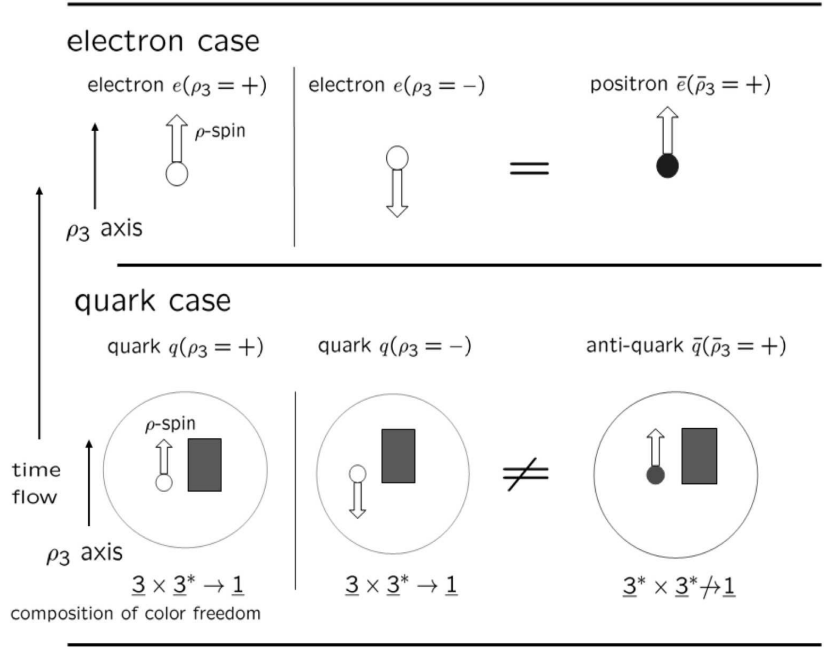}
\caption{Applying Hole Theory for negative-energy solutions
-Free electrons versus Confined-quarks-
Application of H.T. to confined quarks induces violation of 
the color-singlet condition of hadrons. }
\label{fig:1}
\end{center}
\end{figure}
~~~~Applying Hole Theory to respective confined quarks separately 
from the other constituents makes 
the change of quantum numbers explained in Fig.{\ref{fig:1}}, and thus 
violating the color-singlet constraint of Hadrons. \\
~~~~Accordingly the solution of Dirac Eq. is expanded respectively as follows:
\beq
\psi_{D,\alpha}(X)=\sum_{{\bf P},P_{0}=|E|} (u_{\alpha}({\bf P},+|E|)e^{i{\bf P}X-iEt}
+ u_{\alpha}(-{\bf P},-|E|)e^{-i{\bf P}X+iEt}  ), 
\eeq
\beq
u_{\alpha}^{(e^{-})}(-{\bf P},-|E|) \Rightarrow 
v_{\alpha}^{(e^{+})}({\bf P},|E|) \ \ \ \ {\rm for \ electrons}\nonumber
\eeq
\beq
u_{\alpha}^{(a)}(-{\bf P},-|E|) \neq 
v_{\alpha}^{(a^{*})}({\bf P},|E|) \ \ \ \ {\rm for \ quarks} \ \ \ \ \; \nonumber
\eeq
where the $(-)$-frequency solution $u_{\alpha} (-P)$ is 
replaced, in the electron case, by the $(+)$-frequency positron spinor 
$v_{\alpha}(P)$ ; while this is not possible for the confined quark as explained above. 
In the case of confined quarks both of $u_{+,\alpha}(P)\equiv u_{\alpha}({\bf P},|E|)$ 
and 
$u_{-,\alpha}(P)\equiv u_{\alpha}(-{\bf P},-|E|)$ are required for completeness, 
becoming basic vectors of the $SU(2)_{\rho}$ space. 
%%  NO6   %%%%%%%%%%%%%%%%%%%%
\vspace{0.6cm}
\bfl
({\it Overlooked Freedom of $SU(2)_{\rho}$ for Confined Quarks and Chiral States}) \\
\efl
~~~~In the $\widetilde{U}(12)_{SF}$-classification scheme the spin WF of hadrons are 
supposed to 
transform like 
tensors in $\widetilde{U}(4)_{D.S.}$-space 
for the Lorentz transformation 
of CM coordinate of hadrons as 
\beq
\Phi_{\alpha_{1}\cdots \alpha_{n}}^{\beta_{1}\cdots \beta_{m}}(X, r,\cdots)
\propto \Phi_{\alpha_{1}}(X)\cdots\Phi_{\alpha_{n}}(X){\bar{\Phi}}^{\beta_{1}}(X)
\cdots {\bar{\Phi}}^{\beta_{m}}(X). 
\eeq
Here $\Phi_{\alpha} (X)$ is 
representing the respective tensor-component and 
simulating the transformation property of confined quarks, 
and is called ``Urciton''. 
In this connection, it is required only to satisfy the K.G. eqation, 
because the observable is not quarks but 
hadrons. 
\beq
(\frac{{\partial}^2}{\partial X^{2}}-M^2)\Phi_{\alpha}(X,\cdots)=0.
\eeq
{\it $SU(2)_{\rho}$-freedom} : \ Since our master equation contains the 
second-order time derivative it contains{\footnote{In this connection, note 
that the urciton field are described 
by totally $8(=4\times 2)$-components of 
complex field, while the conventional Dirac field by 4-components.}} 
the two-type of Dirac spinors, which constitute the $SU(2)_{\rho}$ space, as
\footnote{The $(+, -)$ sign of $\Phi_{\pm}(X)$ also corresponds to that of $\rho_{3}$
-eigen values of relevant spinors in the rest frame.} 
\beq
\mbox{\rm Basic vectors; \ \ \ \ } 
\ \Phi_{\alpha}(X) = \ \{ \Phi_{+,\alpha}, \Phi_{-,\alpha} \}, 
\nonumber\eeq
defined as solutions of 
\beq
(\gamma_{\mu}\partial_{\mu}\pm M){\Phi}_{\pm}(X)\equiv 0. 
\label{Dirac}
\eeq
They are evidently related by the chirality transformation{\cite{Watanabe}}
{\footnote{Here it may be worthwhile to note that, as far as Eq.(\ref{Dirac})
concerned, the chirality transfrmation is equivalent to the mass 
reversal{\cite{Sakurai}}.
}}
\beq
\Phi_{\pm}(X)=-\gamma_{5} \Phi_{\mp}(X). 
\eeq
%and called as 
%\beq
%\left[ \begin{array}{cc}
%\Phi_{+,\alpha}(X) \ \mbox{\rm with} \ j^{P}={\frac{1}{2}}^{+}& \mbox{\rm Pauli Urciton}
%\\
%\Phi_{-,\alpha}(X) \ \mbox{\rm with} \ j^{P}={\frac{1}{2}}^{-}& 
%\mbox{\rm Chiral Ur-(ex)citon.} \\
%\end{array} \right.
%\nonumber
%\eeq
%The $\Phi_{-,\alpha}(X)$ \ with exotic quantum number leads to new type of 
%``exotic hadrons'': 

We define {\it Chiral states}/ {\it Chiralons}, whose spin WF is tensors in 
$\widetilde{U}(4)_{D.S.}$ space with at least-one chiral urciton-spinor component, 
while {\it Pauli states}/{\it Paulons} with all Pauli-urciton components.
Chiralons show anyhow some ``Exotic'' properties  out of conventional framework, as 
was mentioned in \S1.\\
 
%% NO7 %%%%%

The freedom for quarks, $SU(2)_{\rho}$, had been overlooked for long time. 
Here we describe \\
\bfl
({\it Physical Meaning of $SU(2)_{\rho}$-Freedom})
\efl
~~~~For space-time reflection of CM-coordinates of hadron the WF is transformed, as\\
\beq
X_{\mu}\to X_{\mu}^{'} \ \ : \ \ \Phi(X,r,\cdots)\to \Phi^{'}(X^{'},r,\cdots)
=S^{(H)} \ \Phi (X,r,\cdots) , 
\eeq
where the relevant operator $S^{(H)}$ is given through that for respective 
constituent-urcitons $S$, defined by  

\begin{tabular}{llll}
{\it Space R. }& 
{ $X_{i}^{'} =- X_{i}$ }&
{$S=\gamma_{4}=\rho_{3}\otimes \sigma_{0}$}&
{\it Intrinsic Parity} \\
{\it Time R.} &
{ $X_{0}^{'} =- X_{0}$ }&
{ $S=\gamma_{1}\gamma_{2}\gamma_{3}=i\rho_{2}\otimes \sigma_{0}$}&
{\it Temporal Parity} \\
{\it Space-time R. }&
{ $X_{\mu}^{'} =- X_{\mu}$ }&
{ $S=\gamma_{1}\gamma_{2}\gamma_{3}\gamma_{4}
=-\rho_{1}\otimes \sigma_{0}$}&
{\it Chirality}~. 
\nonumber
\end{tabular}
\vspace{0.3cm}
\\
Note that the chirality transformation is different from the conventional 
chiral transformation, as is compared in Table {\ref{chi}}. However, it is notable that 
the chirality transformation corresponds, mathematically, to the $U(1)$-chiral 
transformation $S={\exp}(i\omega \gamma_{5})$ with an ``angle'' $\omega
=\frac{\pi}{2}$.
\begin{table}[t]
\begin{center}
\caption{Comparison between chirality and chiral transformation}
{\large
\begin{tabular}{cc}
\hline
\hline
Chirality Transf. & Chiral Transf. \\
\hline
discrete & continuous \\
for massive constituent-quarks & for massless current-quarks\\
\hline
\end{tabular}
}
\label{chi}
\end{center}
\end{table}

\newpage
\bfl
({\it A New Symmetry, $SU(2)_{\rho}$, for Hadrons with Light-Quarks}) \\
\efl

The $SU(2)_{\rho}$ freedom for confined quarks leads to a new symmetry in 
hadron physics as follows:\\
~~~~In QCD(, underlying bases for hadron physics,) is valid Chiral Symmetry 
on light quarks in the limit of no effects 
of vacuum condensation. Correspondingly, in hadron physics, simulating 
the constituent-quarks with the urcitons and substituting the transformation 
\fbox{$\gamma_{5}=-\rho_{1}$ } on urcitons for the $\gamma_{5}$-transformation 
on quarks, 
leads to approximate Chiral Symmetry on hadrons with constituent light-quarks. 
This implies also validity of, in the case without direct effects of vacuum 
condensation, 

{\bf  A New $SU(2)_{\rho}$-Symmetry for Hadrons concerning 
Light constituent (even massive) Quarks with respective flavors.} 
\vspace{.6cm}
%%%  NO8  %%%%%%%%%%%%%%%%%%%%%%%%%%%%%%%%%%%
\section{Essentials of Description{\cite{U12}} 
of Composite Hadrons in $\widetilde{U}(12)$-Scheme}
%%%%%%%%%%%%%%%%%%%%%%%%%%%%%%%%%%%%%%%%%%%%% 

In this section we give a brief summary of the relevant 
manifestly-covarinat scheme. 
Note first that, we are not treating a dynamical composite problem, but giving a 
kinematical framework. 

\subsection{General Framework}

\vspace{1em}

{\it Hadron Wave Function} \ \ should represent all {\it attributes} of hadrons, 
such as 
``definite-mass,-$J^{P}$,-Lorentz Transformation Property, and -quark structure''. 

Concerning the definite-Lorentz T.P. and -quark structure, 
we set up the WF $\Phi$ and its Pauli-conjugate $\bar{\Phi}$, 
as follows. 
\beq
\Phi_{A_{1}\cdots}{}^{B_{1}\cdots}(x_{1},\cdots,y_{1},\cdots)&\approx& \psi_{A_{1}}(x_{1})
\cdots \bar{\psi}^{B_{1}}(y_{1})\cdots, \\
\bar{\Phi}^{A_{1}\cdots}{}_{B_{1}\cdots}(x_{1},\cdots,y_{1},\cdots)&\approx& 
\bar{\psi}^{A_{1}}(x_{1})
\cdots \psi_{B_{1}}(y_{1})\cdots, 
\eeq
where $x_{i}$ and $y_{i}$ denote the space-time coordinates of constituent quarks 
and anti-quarks; $A = (\alpha, a)$ and $B= (\beta, b)$ do their Dirac and flavor indices. 
The WF transform like tensors 
in $\widetilde{U}(12)_{SF} \otimes O(3,1)_{\rm Lorentz}$ space. 

\vspace{1em}

{\it Basic Equation} \ \ is, concering the definite-mass, given{\cite{Yukawa}} by 
Yukawa-type Klein-Gordon{\footnote{Note that hadron is observable but quark is not, 
as was mentioned in \S 2.
}} Equation:
\beq
[(\frac{\partial}{\partial X_{\mu}})^{2} - {\cal M}^2 
(r_{\mu}, \partial/\partial r_{\mu})] \Phi (X, r, \cdots)= 0. 
\eeq

Then WF $\Phi$ and its Pauli conjugate $\bar{\Phi}$ are 
expanded by the four-dimensional Fourier amplitudes: 

\beq
\Phi (X,r,\cdots) &=&  \sum_{N, P_{N,\mu}(P_{N,0}>0)}
\{ e^{iP_{N}\cdot X} {\Psi}_{H}^{(+)} (P_{N},r,\cdots)+
e^{-iP_{N}\cdot X} {\Psi}_{\bar{H}}^{(-)}  (P_{N},r,\cdots) \} \nonumber \\
\bar{\Phi} (X,r,\cdots) &=&  \sum_{N, P_{N,\mu}(P_{N,0}>0)}
\{ e^{-iP_{N}\cdot X}  {\bar{\Psi}}_{H}^{(-)} (P_{N},r,\cdots)+
e^{+iP_{N}\cdot X} {\bar{\Psi}}_{\bar{H}}^{(+)} (P_{N},r,\cdots) \} , \nonumber \\
\eeq
where
{\footnote{Note that this implies that $\bar{\Psi}^{(\pm)} (P_{N},r,\cdots)
\equiv \gamma_{4}\cdots[\Psi^{(\mp)}(P_{N},r,\cdots)]^{\dagger}\gamma_{4}\cdots $.}}
\beq
\bar{\Phi}(X, r, \cdots) \equiv \gamma_{4}\cdots \Phi^{\dagger}(X, r, \cdots)
\gamma_{4}\cdots ,\nonumber
\eeq

The description of internal excitation becomes possible due to 
the squared-mass operator on relative coordinates as 

\beq
{\cal M}^2 (r_{\mu}, \partial/\partial r_{\mu}) \Psi_{N}^{(\pm)} (P_{N},r,\cdots)
=M_{N}^2 \Psi_{N}^{(\pm)} (P_{N},r,\cdots), \nonumber
\eeq
\beq
{\cal M}^2 = {\cal M}_{0}^2 + \delta {\cal M}^2. 
\eeq
where the ${\cal M}_{0}^2$ is $U(12)_{SF}$ symmetric and $\delta {\cal M}^2$ represents 
effects of vacuum condensation and of perturvative QCD. 

\vspace{1em} 

{\it Second Quantization} \ \ \ 
The positive / negative frequency parts of WF
$\Phi(X, r,\cdots)$ and  $\bar{\Phi}(X, r,\cdots)$ are supposed to become, as 
a result of the second-quantization, 
\beq
\Psi_{H}^{(+)}(P\cdots) &/& \bar{\Psi}_{H}^{(-)}(P\cdots) ; \ 
\mbox{\rm annihilation-/creation-operator of the relevant hadrons}
\nonumber \\
{\Psi}_{\bar{H}}^{(-)}(P\cdots) &/& \bar{\Psi}_{\bar{H}}^{(+)}(P\cdots) ; \ 
\mbox{\rm creation-/annihilation-operator of their charge-conjugates} \nonumber\\
\eeq
where 
$\bar{\Psi}^{(\pm)} (P)\equiv \gamma_{4}\cdots [\Psi^{(\mp)}(P)]^{\dagger}
\gamma_{4} \cdots$  is the Pauli adjoint both on each lower and upper sufix of 
$\Psi (P\cdots)_{\alpha_{1}\cdots}
{}^{\beta_{1}\cdots}$. 

This leads to {\it the crossing relation of 
relevant hadrons}, which is another attribute of hadrons. \\

%%%  NO9   %%%%%%%%

{\it Composition of WF with definite ${\bf J}={\bf L}+{\bf S}$} \ \ \ \ 
Concerning the attribute of definite $J^{P}$-property; the Fourier amplitudes of WF 
$\Phi(X, r\cdots)$ 
is composed, as  

\beq
\Psi_{J\alpha\cdots}^{(\pm)\beta\cdots}(P,r\cdots)
=\sum_{i,j} 
c_{ij}^{J} \underbrace{W^{(\pm)~(i),\beta\cdots}_{\alpha\cdots} (P) }_{{\rm Spin \ WF}}
\times \underbrace{O^{(j)} (P,r\cdots)}_{{\rm Space-time \ WF}} 
\label{3-6}
\eeq
from complete sets of eigen functions in respective spaces, of the 
$\widetilde{U}(4)_{DS}$ and of the 
relative space-time coordinates. \\
~~~~In Eq.(\ref{3-6}) the spin WF are tensors in $\widetilde{U}(4)_{DS}$ and defined 
by the basic vectors{\footnote{
Basic vectors in $\widetilde{U} (4)_{DS}$ tensor space are given by 
$W_{\alpha}(P)=\{ u_{\alpha} (\mbox{\boldmath $ v$}), 
v_{\alpha}(\mbox{\boldmath $ v$})) \} $, and 
$\bar{W}^{\beta}(P)=\{ \bar{u}^{\beta} (\mbox{\boldmath $ v$})), 
\bar{v}^{\beta}(\mbox{\boldmath $ v$}) \}$ 
which are urciton-Dirac spinors represented 
by the four-velocity of relevant hadrons, $v_{\mu}\equiv P_{\mu}/M (M>0)$. Here
note the restriction, $P_{0}>0$, in Eq.(\ref{38}). This, in addition  to the mass-shell 
condition $P_{\mu}^{2}=-M^2$, implies the zero-th component $v_{0}(\equiv -i v_{4})$ 
becomes $v_{0}=1$ in the rest frame.
}} 
$W_{\alpha}$ and $\bar{W}^{\beta}$ as, 

\beq
W_{\alpha_{1}\cdots\alpha_{n}}^{(+)\beta_{1}\cdots\beta_{m}} (P)=
u_{\alpha_{1}}(P)\cdots u_{\alpha_{n}}(P)\bar{v}^{\beta_{1}}(P)
\cdots\bar{v}^{\beta_{m}}(P) \ \ (P_{0}>0)\nonumber\\
W_{\alpha_{1}\cdots\alpha_{n}}^{(-)\beta_{1}\cdots\beta_{m}} (P)=
v_{\alpha_{1}}(P)\cdots v_{\alpha_{n}}(P)\bar{u}^{\beta_{1}}(P)
\cdots\bar{u}^{\beta_{m}}(P) \ \ (P_{0}>0) , 
\label{38}
\eeq
where $W_{\alpha_{1}\cdots}^{(\pm)\beta_{1}\cdots} (P)$ are Bargmann-Wigner 
spinors, and $u_{\alpha} (P) / v_{\alpha} (P)$ are Dirac spinors for $q/\bar{q}$. 

The spin WF of the Pauli-conjugate $\bar{\Phi}(X, r\cdots)$ are given as 
$\bar{W}_{\beta_{1}\cdots \beta_{m}}^{(\pm)\alpha_{1}\cdots\alpha_{n}} (P) 
= W_{\alpha_{1}\cdots \alpha_{n}}^{(\mp)\beta_{1}\cdots\beta_{m}} (P)$.

In Eq.(\ref{3-6}) the space-time WF, $O(P, r)$, satisfies 
the subsidiary constraint{\cite{PTP}} on 
relative-time, as 

\beq
\la P_{\mu} r_{\mu} \ra =\la P_{\mu} p_{\mu} \ra=0 \stackrel{({\bf P}=0)} {\Rightarrow}
\la t \ra =0 . 
\eeq

Then the 4-dimensional oscillator 
function becomes three-dimensional, effectively, in the hadron rest 
frame as 

\beq
\underbrace{O(P_{\mu},r_{\mu})}_{{\rm tensors \ in \ O(3,1)_{Lorentz}}}
\stackrel{({\bf P}=0)} {\Rightarrow}
\underbrace{O(M_{n}, {\bf r})}_{\rm tensors \ in \ O(3)_{L}}.
\nonumber
\eeq

%%%% NO10 %%%%%%%%%%%%%%%%

\vspace{2em}
\subsection{Formulas concerning basic Vectors in $\widetilde{U}(4)_{DS}$-Tensor Space}

\vspace{1em}

As was mentioned in \S2, our urciton spinors, simulating 
the Lorentz transformation property of confined quarks, are 
required only to satisfy Klein-Goldon equation with 
second-order time-derivative. This leads to existence of two kinds of Dirac 
spinors as its solution. In the following we collect mathematical formulas 
related to the $\widetilde{U}(4)_{DS}$-space. 
\beq
{\rm{\it urcitons }} \ \ \ \ \ \ \ \ \ \ \ 
\Phi_{\cdots\alpha\cdots}^{\cdots}(X;r,\cdots) &\propto&
 \Phi_{\alpha}(X) \ \ \ \ (\alpha : {\rm Dirac \ spinor \ index}). \\
\widetilde{U}(4)_{DS}&\stackrel{{\bf P}=0}{\Rightarrow}
&U(4)_{S} \supset SU(2)_{\sigma}\times 
SU(2)_{\rho}\nonumber
\eeq

{\it K.G.Eq.}
\beq
&&(\Box - M^2)\Phi_{\alpha}(X) = \left[ (\gamma_{\mu}\d_{\mu} + M)
(\gamma_{\lambda}\d_{\lambda}-M)\Phi
\right] _{\alpha} =0
\nonumber \\
&&\Phi_{\alpha}(X)= \Phi_{+,\alpha}(X)+ \Phi_{-,\alpha}(X)  \ \ \ \ \nonumber\\
&&(\gamma_{\mu} \d_{\mu} \pm M) \Phi_{\pm}(X)\equiv 0  \ \ (M>0) .
\eeq

{\it Basic Vectors for $SU(2)_{\rho}$}  \ \ :\ \ \ 
$\Phi_{r}(X) = 
\{ \Phi_{+}(X) ,\Phi_{-}(X) \}$. 

\vspace{1em}

{\it Fourier Expansion of urciton spinor}
\beq
\Phi_{\alpha}(X)= \sum_{{\bf P}(P_{0}\equiv E_{P} >0), r,s,\bar{r},\bar{s}=\pm1} 
(b_{r, s, {\bf P}} u_{r,s,\alpha}(P) e^{iPX}
+d^{\dagger}_{\bar{r}, \bar{s}, {\bf P}} v_{\bar{r},\bar{s},\alpha}(P) e^{-iPX})
\nonumber\\
\bar{\Phi}^{\alpha}(X)= \sum_{{\bf P}(P_{0}\equiv E_{P} >0), r,s,\bar{r},\bar{s}=\pm1} 
(b^{\dagger}_{r, s, {\bf P}} \bar{u}_{r,s}^{\alpha}(P) e^{-iPX}
+d_{\bar{r}, \bar{s}, {\bf P}} \bar{v}_{\bar{r},\bar{s}}^{\alpha}(P) e^{+iPX}).
\nonumber\\
\eeq
Here it should be noted that a summation on the new freedom, $SU(2)_{\rho}$, 
denoted as $r=\pm$ 
(the eigen value of $\rho_{3}$-spin), is appearing in additon to 
the conventional one $s=\pm$ on $SU(2)_{\sigma}$.\\

{\it Fourier Conjugate-Basic Vectors in $\widetilde{U}(4)_{DS}$-Tensor 
Space}
\beq
W_{\alpha}(v)&=& \{ u_{\alpha}(v), v_{\alpha}(v) \}, \nonumber\\
\bar{W}^{\beta}(v)&=&\{ \bar{u}^{\beta}(v), \bar{v}^{\beta}(v) \}. \ \ \ \ 
(v_{\mu}\equiv \frac{P_{\mu}}{M} \ ; \ v_{0}> 0)
\eeq

\vspace{1em}

{\it Complete set of basic vectors in $SU(2)_{\rho}$-
space}
\beq
u_{r}&=&\{ u_{+}, u_{-} \} , \ 
v_{r}=\{ v_{+}, v_{-} \} \nonumber\\
\bar{u}_{r}&=&\{ \bar{u}_{+}, u_{-}{} \} , \ 
\bar{v}_{r}=\{ \bar{v}_{+}{}, \bar{v}_{-}{} \}.  
\eeq

{\it Chirality Partners}
\beq
&&W_{\pm}(v)=-\gamma_{5} W_{\mp} (v) \ ; \ \bar{W}_{\pm}=\bar{W}_{\mp}(v) \gamma_{5}
\\
&&\left\{
\begin{array}{l}
r=+ ; \ \ \ \mbox{\rm Pauli- (urciton) spinor }\\
r=- ; \ \ \ \mbox{\rm Chiral- (urciton) spinor }
\end{array}
\ \ \ \mbox{\rm opposite relative-parity}
\right.\nonumber
\eeq
\\
%%%% NO11 %%%%%%%%%%%%%
\subsection{Static-$U(4)_{S}$-embedded spinor WF}

In order to embed{\footnote{
As for details, see Ref.~{\citen{Ds}}}} the static $U(4)_{S}$-symmetry in the covariant 
$\widetilde{U}(4)_{D.S.}$-space, it is necessary to replace the basic vectors 
$\bar{W}^{\beta}(v)$ by $\bar{W}^{\beta}_{U}(v)$ and to make corresponding 
modification on the spinor WF of general hadrons, as follows:\\

(basic vectors)
\beq
{W}_{\alpha}(P) &\Rightarrow& {W}_{U, \alpha}(P) \equiv 
{W}_{\alpha} (P) \nonumber\\
\bar{W}^{\beta}(P) &\Rightarrow& \bar{W}_{U}^{\beta}(P) \equiv 
\left[ \bar{W} (P) F_{U} (v)\right]^{\beta}~
( F_{U}(v) \equiv -iv\cdot \gamma \stackrel{({\bf P}=0)} {\Rightarrow} \gamma_{4} )
\nonumber\\
\bar{W}_{U}^{\beta}(P)&\stackrel{({\bf P}=0)} {\Rightarrow}&
\left[ {W}^{\dagger} ({\bf P}=0) \gamma_{4}\gamma_{4} \right]^{\beta}
=W^{\dagger,\beta}({\bf P}=0) \leftarrow \mbox{\rm Hermite Conjug.}
\eeq

The $F_{U}(v)$, unitarizer, becomes $\gamma_{4}$ in the rest frame of relevant hadrons, 
while Pauli-adjoint does Hermite conjugates. \\
~~~~Accordingly the new scalar product of basic vectors becomes equal to 
the unitary-invariant product in the rest frame of relevant hadrons, as 
\beq
\la \bar{W}_{U} (P)^{\alpha} W_{U}(P)_{\alpha} \ra \stackrel{({\bf P}=0)} {\Rightarrow}
\la W^{\dagger} (P)^{\alpha} W (P)_{\alpha} \ra _{{\bf P}=0} .\\
\nonumber
\eeq

(Spinor WF of Hadron)\\
\beq
W_{\alpha_{1}\cdots\alpha_{n}}{}^{\beta_{1}\cdots\beta_{m}}(P)
&=&W_{\alpha_{1}}(P)\cdots W_{\alpha_{n}}(P)
\bar{W}^{\beta_{1}}(P)\cdots \bar{W}^{\beta_{m}}(P)\nonumber\\
&\downarrow&\nonumber\\
W_{U, \alpha_{1}\cdots\alpha_{n}}{}^{\beta_{1}\cdots\beta_{m}}(P)
&=&W_{U,\alpha_{1}}(P)\cdots W_{U,\alpha_{n}}(P)
\bar{W}_{U}^{\beta_{1}}(P)\cdots \bar{W}_{U}^{\beta_{m}}(P)\nonumber\\
&\stackrel{({\bf P}=0)} {\Rightarrow}&W_{\alpha_{1}}({\bf P}=0)\cdots W_{\alpha_{n}}({\bf P}=0)
{W}^{\dagger, \beta_{1}}({\bf P}=0)\cdots W^{\dagger, \beta_{m}}({\bf P}=0).\nonumber\\
\eeq
~~~~Accordingly the scalar product of the $U(4)_{S}$-embedded spin WF becomes equal to 
the unitary-invariant product in the rest frame, as 
\beq
\la \bar{W}_{U} (P)^{\alpha_{1}\cdots\alpha_{n}}_{U, \beta_{1}\cdots\beta_{m}}
 &&W_{U}(P)_{\alpha_{1}\cdots\alpha_{n}}^{\beta_{1}\cdots\beta_{m}} \ra \nonumber\\
 &&\stackrel{({\bf P}=0)} {\Rightarrow}
(\prod_{i=1}^{n} \la W^{\dagger, \alpha_{i}} (P) W_{\alpha_{i}} (P) \ra _{{\bf P}=0})
(\prod_{j=1}^{m} \la W^{\dagger, \beta_{j}} (P) W_{\beta_{j}} (P) \ra _{{\bf P}=0})
 .\nonumber\\
\eeq

%%%%%%%%%%%%%%%%%%%%%%%%%%%%%%%%%%
\section{Revisal of Meson and Baryon Wave Functions and Chiral States}

In this section we give a brief review on the new treatment of the conventional 
($q\bar{q}$)-mesons and ($qqq$)-baryons in the $\widetilde{U}(12)$-scheme, and decsribe the 
level structures, referring to the chiral states. 

%%%%  NO12  %%%%%%%%%%%%

\subsection{Spinor WF of $(q\bar{q})$-Mesons and $(qqq)$-Baryons }

In this sub-section we give the concrete form of spinor WF of $(q\bar{q})$ 
mesons and 
$(qqq)$ baryons.\\

{\it Dirac Spinor }
\beq
\psi_{D}(X)_{\alpha}= \Phi_{+,\alpha}(X)=
\sum_{{\bf P}(P_{0} >0)} 
(W_{q, +, \alpha}(P) e^{iPX}
+W_{\bar{q}, +, \alpha}(P) e^{-iPX}).
\eeq

{\it Urciton Spinor } $\Phi_{\alpha}(X)=\Phi_{+,\alpha}(X)+\Phi_{-,\alpha}(X)$
\beq
\Phi_{\alpha}(X)&=&
\sum_{{\bf P}(P_{0}>0), r,\bar{r}=\pm}
(W_{q, r, \alpha}(P) e^{iPX}
+W_{\bar{q}, \bar{r}, \alpha}(P) e^{-iPX})
\nonumber\\
\bar{\Phi}^{\beta}(X)&=&
\sum_{{\bf P}(P_{0}>0),  r,\bar{r}=\pm} 
(\bar{W}_{q, r}{}^{ \beta}(P) e^{iPX}
+\bar{W}_{\bar{q}, \bar{r}}{}^{ \beta}(P) e^{-iPX}).
\eeq

{\it Complete set of Basic Vectors in $SU(2)_{\rho}$ space}
\beq
W_{\alpha}(P)&=&\{ W_{q,r,\alpha}(P), W_{\bar{q}, \bar{r}, \alpha}(P) \}, r,\bar{r} =\pm \nonumber\\
\bar{W}^{\beta}(P)&=&
\{ \bar{W}_{q,r}^{\beta}(P), \bar{W}_{\bar{q}, \bar{r}}^{\beta}(P) \}, r,\bar{r} =\pm ~~.
\eeq

{\it Meson Spinor} \ \  \  bi-Dirac Spinor
\beq
W_{\alpha}{}^{\beta}(P)=W_{q, r, \alpha}(P) \bar{W}_{\bar{q}, \bar{r}}^{\beta}(P)
\eeq
\beq
(r, \bar{r})=&& \ \ (+, +) \ \ \ \mbox{\rm boosted Pauli States}\nonumber\\
&&\left.\begin{array}{c}(+, -)\\(-, +)\\(-, -)\end{array} \right\} 
\mbox{\rm``Chiral States''~~~~.}
\nonumber 
\eeq

{\it Baryon Spinor} \ \  \  tri-Dirac Spinor
\beq
W^{(B)}_{\alpha\beta\gamma}(P)=W_{q,r_{1}, \alpha}(P) W_{q,r_{2}, \beta}(P) 
W_{q,r_{3}, \gamma}(P) : 
\mbox{\rm for Baryons}
\eeq
\beq
(r_{1}, r_{2}, r_{3})
=&& \ \ (+, +, +) \ \ \ \mbox{\rm boosted Pauli States}\nonumber\\
&&\left.\begin{array}{c}(+, +, -)\\(+, -, -)\end{array} \right\} 
\mbox{\rm``Chiral States''}
\nonumber
\eeq
\beq
W^{(\bar{B})}_{\alpha\beta\gamma}(P)=
W^{(B)}\{ W_{q,r_{i}, \alpha}\Rightarrow W_{\bar{q},\bar{r}_{i},\alpha} \} 
: \mbox{\rm for anti-Baryons}~~.
\nonumber
\eeq

%%%  NO13   %%%%%%%%%%%%%%%%
\vspace{2em}

\subsection{Level Structure and Excitation 
Trajectories of Mesons}

(Ground Sates) in $\widetilde{U}(12)_{SF}$-scheme are classified, respectively, into 
the multiplets as follows: 

Meson: $(\underline{12} \times  \underline{12}^{*})=\underline{144}$
\let\tabularsize\scriptsize
\begin{table}[h]
\begin{center}
\begin{tabular}{lcc|cccccc|}
&&\multicolumn{6}{r}{(Chiral States)~~~~~~~}\\ 
\cline{4-9}
&$P_{s}^{(N)}$&$V_{\mu}^{(N)}$&$P_{s}^{(E)}
$&$V_{\mu}^{(E)}$&$S^{(N)}$&$A_{\mu}^{(N)}$&S$^{(E)}$&$A_{\mu}^{(E)}$\\
$J^{PC}$&$0^{-+}$&$1^{--}$&$0^{-+}$&$1^{--}$&$0^{++}$&$1^{++}$&$0^{+-}$&$1^{+-}$\\
\cline{4-9}
\end{tabular}
\end{center}
\end{table}
\\
Baryon: $(\underline{12} \times \underline{12}\times \underline{12})_{\rm Symm}
=\underline{364}=\underline{182}_{B} \oplus \underline{182}_{\bar{B}}$
\\
\begin{table}[h]
\begin{center}
\begin{tabular}{l|ccccc|}
\multicolumn{2}{r}{$\underline{56}$~} & $\Delta_{3/2}^{\oplus}$ & $N_{1/2}^{\oplus}$ &&
\multicolumn{1}{c}{}\\
\cline{2-6}
$\underline{182} \Rightarrow $
&$\underline{70}$& $\Delta_{1/2}^{\ominus}$ & $N_{3/2}^{\ominus}$ &
$N_{1/2}^{\ominus}$ & $\Lambda_{1/2}^{\ominus}$\\
&$\underline{56^{'}}$& $\Delta_{3/2}^{\oplus}$ & $N_{1/2}^{\oplus}$ &
 & \\
\cline{2-6}
\multicolumn{6}{c}{~~~~~~~(Chiral States)}
\end{tabular}
\end{center}
\end{table}

\vspace{1em}

(Excited States ) in $\widetilde{U}(12)_{SF}\otimes O(3)_{L} $-scheme show{\cite{R}}, respectively, 
the excitation trajectories, as 
\beq
\mbox{\rm (mass)}^{2} \ \ \ M_{N}^2 = M_{0}^2 +N\Omega
\label{excite}
\eeq
\beq
\mbox{\rm meson: } \ M_{M}^{(N)}&=&m_{q}^{(N)} + m_{\bar{q}}^{(N)}\nonumber\\
\mbox{\rm baryon: } \ M_{B}^{(N)}&=&m_{q_{1}}^{(N)}+m_{q_{2}}^{(N)}+m_{q_{3}}^{(N)},\nonumber
\eeq
where $m_{q}^{(N)}$ being the mass of quarks in N-th excited states. 
In Eq.(\ref{excite}) the mass of ground states $M^{(G)}=M_{0}$ is given. 
In the ideal case, as a sum of that of respective constituent-quarks $m_{q}$, 
taking $m_{q}^{(0)}=m_{q}$, while the mass of N-th excited states $M^{(N)}$ 
are given as a sum of those of respective excited-constituent quarks $m_{q}^{(N)}$, 
being given by $m_{q}^{(N)}=\gamma_{N} m_{q}$ with $\gamma_{N} \equiv M^{(N)}/ M^{(G)}$. \\

\let\tabularsize\footnotesize

(Effective Chiral Symmetry) We suppose a phenomenological 
rule for approximate chiral symmetry; 
\beq
m_{q}^{(N)2} \ll \Lambda_{conf}^{2} \approx 1 {\rm GeV}^2 . 
\label{nn}
\eeq

This leads to a conditon on number of excitation quanta as 

\begin{table}[h]
\begin{center}
\begin{tabular}{|ccc|}
\hline
$(n\bar{n})_{\rm Meson}$&$(n\bar{c})_{\rm Meson}$&$(n\bar{b})_{\rm Meson}$ \\
$N\leq 1 \ {\rm or } \  2$&  $N\leq 0 \ {\rm or } \ 1$ & $N\leq 0 \ {\rm or } \ 1$\\
\hline
\end{tabular}
\end{center}
\end{table}

{\hspace{-2em}}Accordingly, for the Lower-mass states with these $N$ values we 
expect{\cite{U12,protovino}} {\it the Existence of Chiral States}.
The situations on the meson excitation-trajectory are shown in Fig.2. 
\let\tabularsize\normalsize

%%%  NO14   %%%%%%%%%%%%%%%%%%%%%

\begin{figure}[h]
\begin{center}
\includegraphics[width=13cm,height=13cm]{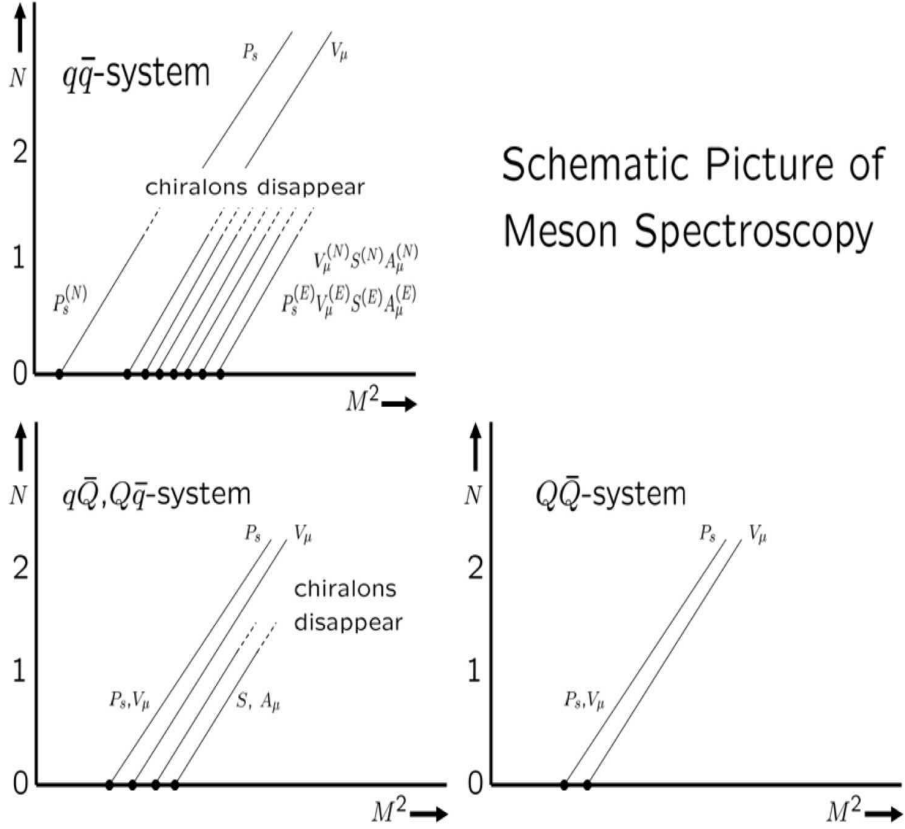}
\label{fig:R1}
\end{center}
\caption{Schematic Picture of Meson Spectroscopy. \ \  
For the lower-mass states, in the L-L and H-L system, with the excitation number $N$, 
given below Eq.(\ref{nn}), are expected the existence of chiralons, whose 
trajectries will 
disappear in the higher mass region.
}
\end{figure}

%%%  NO15  %%%%%%%%%%%%
%%%  NO16  %%%%%%%%%%%%%%%%%%%%%%%%%%
\section{Multi-Quark System 
and Near-Threshold Resonances }
%%%%%%%%%%%%%%%%%%%%%%%%%%%%%%%%%%%%%%%%%%%%% 

\begin{figure}[htbp]
\begin{center}
\includegraphics[width=13cm,height=12cm]{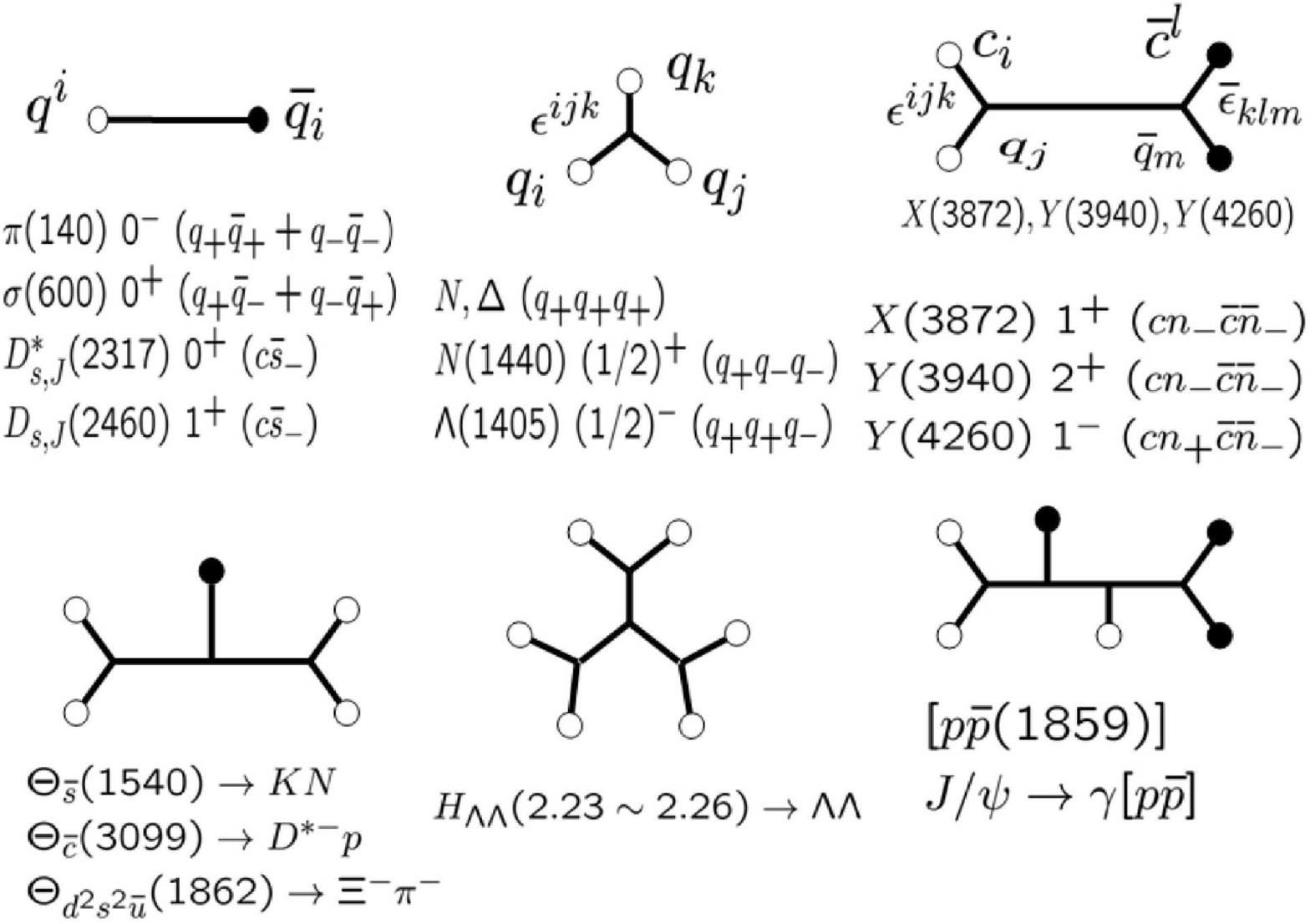}
\caption{Overview of Hadrons in $\widetilde{U} (12)$-Classification Scheme.  
The view is obtained applying the JSQM{\cite{Imachi}} in our covariant classification 
scheme, where the $q_{+}$/$q_{-}$ denotes Pauli/chiral urciton-spinor, repectively. The 
hadrons described with $q_{-}$ or $\bar{q}_{-}$ show anyhow some exotic-properties. 
}
\label{fig:Overview}
\end{center}
\end{figure}

An overview of general multi-quark hadrons, obtained by applying the Joined Spring 
Quark Model{\cite{Imachi}}, 
in the $\widetilde{U}(12)_{SF}$-classification scheme 
is given in Fig.\ref{fig:Overview}. 
The JSQM had been proposed so as to lead to the triality-zero and color-singlet 
multi-quark hadrons. Here the model applied in the $\widetilde{U}(12)$-scheme 
is manifestly 
covariant, and gives 
chirality($\gamma_{5}$)-symmetric (concerning the light quarks) mass 
spectra in the ideal limit. First in this section I shall 
give an interpretation of basic properties of new ``exotic'' hadrons in 
the $\widetilde{U}(12)$ scheme. Then I give some comments in identifying 
the near-threshold resonances observed at BES. \\
%%%%%%%%%%%%%%%%%%%
%%%%%%%%%%%%%%%%%%%%
\subsection{Properties of Tetra-quarks - an Example{\cite{ourX}} of $X(3872)$-Family}
\begin{figure}[t]
\begin{center}
\includegraphics[width=14cm,height=4cm]{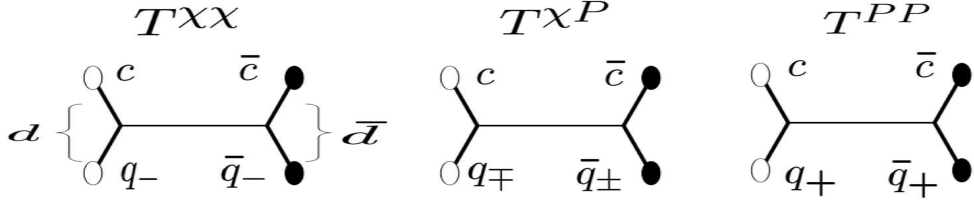}
\caption{Classification of Tetra-quarks. The $SU(2)_{\rho}$ WF
 is classified into the three groups.}
\label{fig:4quark}
\end{center}
\end{figure}

\bfl
({\it Experimental Candidates of Tetra-quarks})
\efl
~~~~The three resonances $X(3872)$, $Y(3940)$, and $Y(4260)$ observed recently 
in Belle and BaBar experiments, may be promising candidates of the tetra-quark system. 
The properties of three resonances are collected in Table {\ref{tab:data}}. 
\let\tabularsize\footnotesize
\begin{table}[h]
\begin{center}
\caption{Experimental Data of $X(3872)$-Family}
\begin{tabular}{ccccc}
\hline
\hline
{\large  }& {\large I}&{\large $J^{PC}$}&{\large Decay Channel}&{\large $\Gamma$(MeV)} \\
\hline
{\large $X(3872)$ }&{\large $0,1$ }&{\large $1^{++}$}&{\large $\omega (\rho)+J/\psi$}&{\large $<2.3$} \\ 
{\large $Y(3943)$ }&{\large $0$ }&{\large $2^{++}$}&{\large $\omega+J/\psi$}&
{\large $\approx 87 $} \\
{\large $Y(4260)$ }&{\large $0$ }&{\large $1^{-}$}&
{\large $S(\sigma/f_{0}(\to \pi^{+}\pi^{-}))J/\psi$}&{\large $50-90 $} \\
\hline
\end{tabular}
\label{tab:data}
\end{center}
\end{table}
%%%%%%%%%%
\bfl
({\it Level Structure})
\efl
~~~~The $SU(2)_{\rho}$-WF of tetra-quark system  in JSQM is 
classified, depending upon the ( chiral or Pauli type ) properties of constituent 
diquak/anti-diquark, into the three types (given in Fig.{\ref{fig:4quark}}), where 
$T^{\chi\chi}$ etc. denotes tetra-quark system with the constituent diquark and 
anti-diquark as follows:
\beq
T^{\chi\chi} &\equiv& [ d^{\chi}(cq_{-})\cdot \bar{d}^{\chi}(\bar{c}\bar{q}_{-}) ], \ 
T^{\chi P}\equiv [ d^{\chi}(cq_{-})\cdot \bar{d}^{P}(\bar{c}\bar{q}_{+}) ]~/~ 
[ d^{P}(cq_{+})\cdot \bar{d}^{\chi}(\bar{c}\bar{q}_{-}) ], \nonumber\\
T^{PP}&\equiv& [ d^{P}(cq_{+})\cdot \bar{d}^{P}(\bar{c}\bar{q}_{+}) ].
\eeq

The possible Spin-Parity values $J^{P}$ for constituent di-quarks and tetra-quarks are, 
reppectively, as 
\beq
\mbox{\rm Diquarks;} \ \ \ d^{P}(0^{+}, 1^{+}) , \ \ d^{\chi}(0^{-}, 1^{-})
~~~~~~~~~~~~~~~~~~~~~~~~~~~~~~~~~~~~~~~~~
\eeq
\beq
&&\mbox{\rm Tetraquarks;} \ \ \ T^{\chi\chi}[(0,1,2)^{+}] , 
\ \ T^{\chi P}[(0,1,2)^{-}], \ \ 
T^{PP}[(0,1,2)^{+}].~~~~~~~~~
\eeq
%\beq
%J^{PC} \left\{ 
%\begin{array}{ll}
%(0^{-})\cdot (0^{-}) = 0^{++} &  \nonumber\\
%(0^{-})\cdot (1^{-}) = 1^{+}& ; X(3872) \leftarrow (q=n) \nonumber\\
%(1^{-})\cdot (1^{-}) = 0^{+}, 1^{+}, \underline{2^{+}}&
% ; Y(3943) \leftarrow (q=n), \nonumber
%\end{array} \right.
%\eeq
%%\beq
%J^{PC} \left\{ 
%\begin{array}{ll}
%(0^{-})\cdot (0^{+}) = 0^{-} &  \nonumber\\
%(0^{-})\cdot (1^{+}) = 1^{-}& ; Y(4260) \leftarrow (q=n) \nonumber\\
%(1^{-})\cdot (1^{+}) = 0^{-}, 1^{-}, 2^{-}&, \nonumber
%\end{array} \right.
%\eeq
%
%\beq
%J^{PC} \left\{ 
%\begin{array}{ll}
%(0^{+})\cdot (0^{+}) = 0^{++} &  \nonumber\\
%(0^{+})\cdot (1^{+}) = 1^{+}& \nonumber\\
%(1^{+})\cdot (1^{+}) = 0^{+}, 1^{+}, 2^{+}&, \nonumber
%\end{array} \right.
%\eeq
\bfl
({\it Mass Relation})
\efl
~~~~The mass of tetra-quarks are given by 
\beq
M_{T}&=&M_{T}^{(0)}+(\delta^{\chi} M_{T} + \delta^{J} M_{T}), 
\eeq
where $M_{T}^{(0)}$ is $U(12)$ symmetric ; $\delta^{\chi} M$ 
represents the effect of vacuum condensation 
$\la q \bar{q} \ra_{VEV}$; and $\delta^{J} M_{T}$ does the one due to hyperfine 
spin-spin interaction 
\beq
H_{J} \approx (\bar{c} \sigma_{\mu\nu}c)
(\bar{q}_{U}\sigma_{\mu\nu}q)\propto 
\la {\bf \sigma^{(c)}}\ra \la {\bf \sigma^{(q)}}\ra.
\label{HJ}
\eeq

The $M^{(0)}_{T}$ is given by, and satisfies the following equations 
\beq
M_{T}^{(0)}=\sum_{i(\rm constit.)} m_{i} =2(m_{c} + m_{q}), 
\eeq
\beq
M_{T}^{(0)}=M_{d}^{(0)}(cq)+M_{\bar{d}}^{(0)}(\bar{c}\bar{q})
=M_{\psi}^{(0)}(c\bar{c})+M_{M}^{(0)}(q\bar{q}). 
\label{Mchi}
\eeq

Both the $\delta M^{\chi}_{T}$ and $\delta M^{J}_{T}$ are supposed to be given 
as a sum of those of constituent diquarks. Then their numerical values are able to 
estimate by using the following formulas and the knowledge obtained from the analyses of 
$D_{s}(c\bar{s})$ mesons.
\beq
\mbox{\rm On \ \ $\delta^{\chi} M ; $} ~~~~~
\delta^{\chi} M (\equiv M^{\chi}-M^{P} ) = \delta^{\chi} M_{d}(cq)+
\delta^{\chi} M_{\bar{d}}(\bar{c}\bar{q}); 
\eeq
\beq
\delta^{\chi} M_{d}(cq)= - \delta^{\chi} M_{D} (c\bar{q}), 
\label{a}
\eeq
\beq
\delta^{\chi} M_{D}(c\bar{n})&=&242 \mbox{\rm MeV}, \nonumber\\
\delta^{\chi} M_{D}(c\bar{s})&=&348 \mbox{\rm MeV}. 
\label{b}
\eeq
%(input from $D_{s}(c\bar{s})$, $D_{n}(c\bar{n})$%[$D_{s}(2317:0^+)$, $D_{s}(2460:1^+)$])
%
%\begin{figure}
%\begin{center}
%\includegraphics[width=8cm,height=4cm]{VEV.eps}
%\label{fig:VEV}
%\caption{The figure caption should be here!!}
%\end{center}
%\end{figure}
\beq
\mbox{\rm On \ \ $\delta^{J} M$; } ~~~~~~
\delta^{J} M \equiv M(J=1) - M(J=0) , ~~~~~~~~~~~~~~~~
\label{c}
\eeq
\beq
\delta^{J} M_{d}(cq) =\frac{1}{2} \delta^{J} M_{D}(c\bar{q}) = 71{\rm MeV} . 
\label{c}
\eeq
Here I add several remarks on the equations given above : \\
%The several equations given above are derived from rather general assumptions: \\

The relation (\ref{Mchi}) reflects the static $U(12)$-symmetry in the ideal limit, 
a basic assumption in our classification scheme, and explains one of the features F1 of 
new exotic-hadrons mentioned in \S 1. \\
~~~~The relation (\ref{a}) between chiral splittings 
$\delta^{\chi} M$ of $(cq)$ diquarks 
and of $D(c\bar{q})$ mesons are derived, considering the vacuum condensation effects 
through their constituent light-quark. It is notable that the opposite sign between 
the splittings reflects the charge conjugation property of the c-number scalar bilinear 
$(\bar{\Phi} 1 \Phi)$ of urciton spinors. \\
~~~~The relation (\ref{c}) between hyperfine splittings in the two systems is derived, 
considering the color-gauge invarinat and static $U(4)$-invarinat interaction between 
the heavy $c$-quark and the light $q$-quark, Eq.(\ref{HJ}). 
The same sign between the 
splittings reflects the properties of c-number tensor bilinear 
$(\bar{\Phi} \sigma_{\mu\nu}\Phi)$ (similarly as in Eq.(\ref{a})), and the 
factor $1/2$ represents the difference 
between color-space WF of the two systems. \\
~~~~The numerical values in (\ref{b}) are obtained, 
respectively, on $\delta^{\chi} M_{D}(c\bar{s})$ 
from the experimental value, and on $\delta^{\chi} M_{D}(c\bar{n})=(a/b) 
\delta^{\chi}M_{D}(c\bar{s})$ from 
the experimental value $a/b=1/1.44$, where $a\propto \la n\bar{n}\ra _{\rm VEV}$ 
and $b\propto \la s\bar{s}\ra_{\rm VEV}$. \\
%The numerical value in (\ref{c}) is obtained from the phenomenological analysis of 
%$D(c\bar{s})$ mesons.\\

\begin{figure}[b]
\begin{center}
\includegraphics[width=13cm,height=5cm]{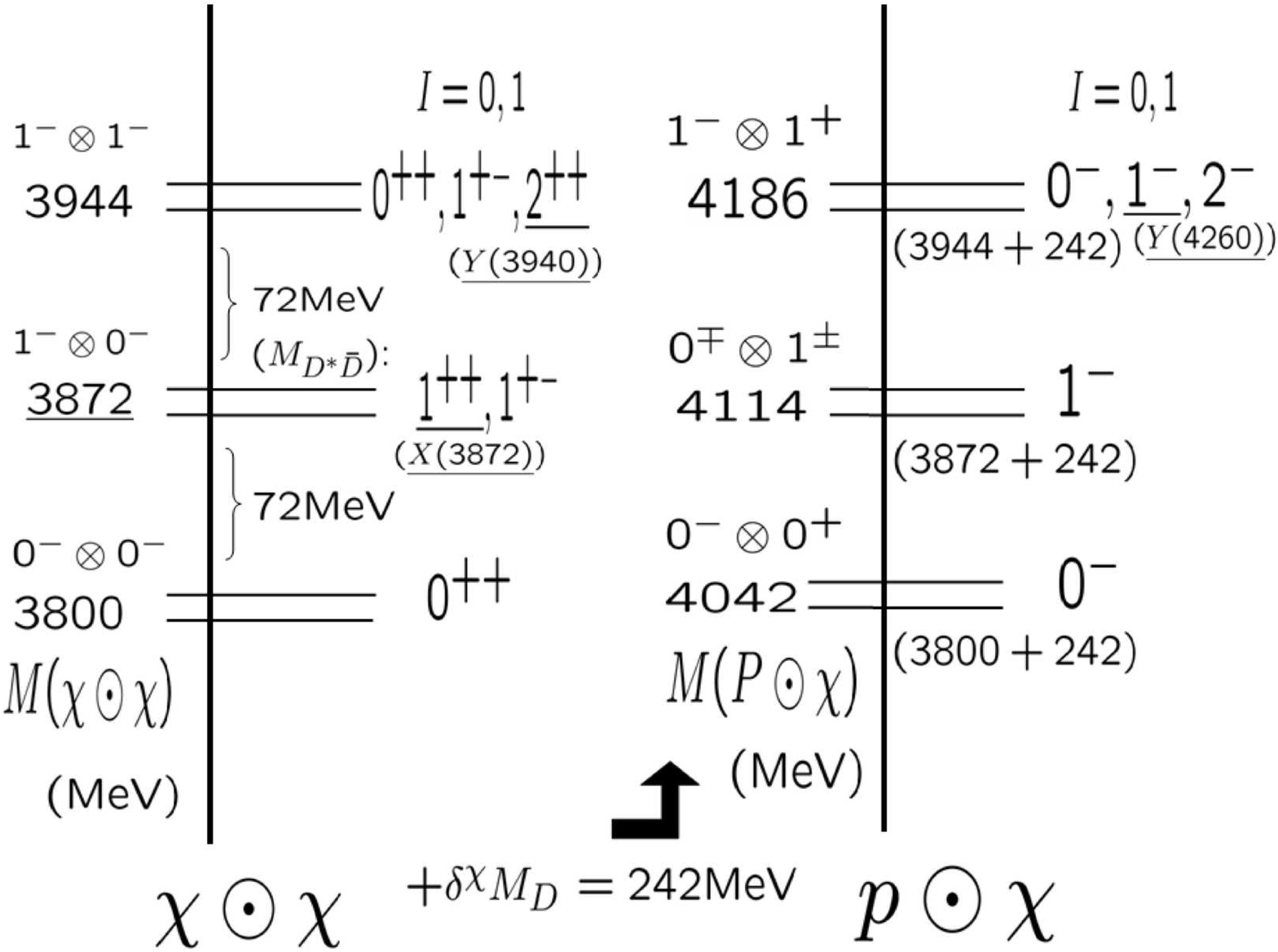}
\caption{Mass Spectra and Level Structure of Ground State $X(3872)$ Meson Family. 
The mass of $X(3872)$ is used as input, and a tentative assignment of the $Y(3940)$ 
and $Y(4260)$ 
is also made.}
\label{fig:mass-spectra}
\end{center}
\end{figure}

\bfl
({\it Mass Spectra})
\efl
~~~~The level structure and mass spectra of ground state $X(3872)$ meson families, 
thus determined, are shown 
in Fig.{\ref{fig:mass-spectra}};
%The mass values with the input of $X(3872)$ 
%of respective members of the 
%$\chi\cdot\chi$ and $\chi\cdot P$ groups in the structure pattern shown above. 
where %only the results for the $T^{\chi\chi}$ and $T^{\chi P}$ states are given. 
the $J^{PC}$-structure of $T^{PP}$ group is identical to that of $T^{\chi \chi}$, 
while the $T^{\chi P}$ group has the same J-structure and opposite Parity 
as the other groups. 
The mass values of corresponding members of these groups are heavier by twice of 
$\delta^{\chi} M_{D}= 242 {\rm MeV}$ in order of $M(\chi\chi)< M(P\chi)<M(PP)$. 
However, the $T^{P P}$ members are expected having too 
large width to be observed and not shown in Figure, see the next sub-sections. \\
%%%  NO17 %%%%%%%%
%%%% NO18  %%%%%%%%%%%%%%%
~~Taking into account these considerations, three resonances 
$X(3872)$, $Y(3940)$, and $Y(4260)$ 
may be assigned as shown in 
Fig.\ref{fig:mass-spectra}. 

\bfl
({\it Decay Mechanism})
\efl
\begin{figure}[htbp]
\begin{center}
\includegraphics[width=13cm,height=4cm]{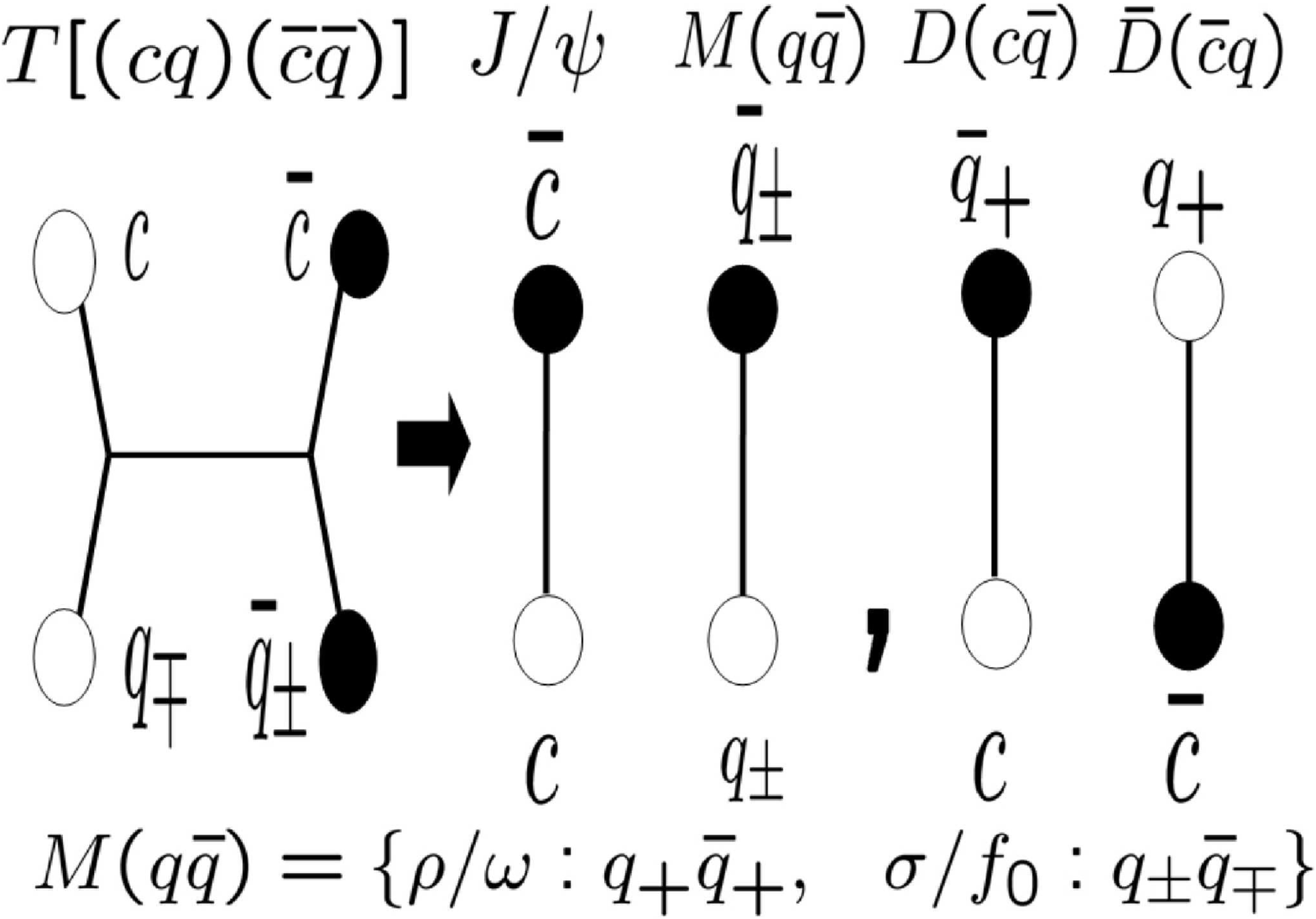}
\caption{Decay of $T[(cq)(\bar{c}\bar{q})]$ into $J/\psi + M(q\bar{q})$, 
$D(c\bar{q})+\bar{D}(\bar{c}q)$. 
In the first decay channel the two cases of $M(q\bar{q})$  
are observed. }
\label{fig:rearrangement}
\end{center}
\end{figure}
~~~~In the relevant close-to-threshold decay, 
the dominat amplitude is considered to come from rearrangement of constituent quarks. 
This process, shown in Fig.{\ref{fig:rearrangement}}, contains 
overlapping of initial and final WF.\\
\bfl
({\it Overlapp of quark spinor WF and ${\it \rho_{3} line rule}$})
\efl
~~~~The overlapping contains the factor concerning the light-quark line, which has the 
properties, as 

(1) at threshold \ \ $(\bar{q}_{U,\pm}(\mbox{\boldmath $v$}),
 {q}_{\pm}(\mbox{\boldmath $v$}))=1$, $(\bar{q}_{U,\pm}(\mbox{\boldmath $v$}),
 {q}_{\mp}(\mbox{\boldmath $v$}))=0$\\

(2) near threshold \ \ $(\bar{q}_{U,\pm}(\mbox{\boldmath $v$}_{F}),
 {q}_{\mp}(\mbox{\boldmath $v$}_{I}=0))=\frac{|\mbox{\boldmath $P$}_{F}|}{2 M_{F}}
 \equiv \epsilon (|\mbox{\boldmath $P$}_{F}|)\ll 1 $.\\

These two relations reflect the basic contents of static $U(4)$ symmetry, as 
{orthonormality} of spinor WF in {\it static $U(4)_{S}$ space} and $\rho_{3}$
(third component of $\rho$-spin) conserveation at threshold. We called them 
the {\it $\rho_{3}$-line rule}.\\

\bfl
({\it Decay Width of Tetra-quarks with ideal $SU(2)_{\rho}$ WF}) 
\efl
~~~~In Table {\ref{tab:TetraDecay}} we have collected 
the qualitative decay properties of T-mesons 
with the typical $SU(2)_{\rho}$ WF, derived from the $\rho_{3}$-line rule, are given. 
\begin{table}
\begin{center}
\caption{Decay Property of Tetra-Quark Mesons}
\begin{tabular}{cccc}
\hline
\hline
Structure of WF &allowed/forbidden &Structure of Decay Channel &Wirth $\gamma$\\
\hline
$T[(cq_{+})(\bar{c}\bar{q}_{+})]$& $\Rightarrow$ (allowed) &
$D(c\bar{q}_{+})+\bar{D}(\bar{c}q_{+})$ & $\Gamma^{0}$ a few GeV \\
&&$\psi(c\bar{c})+ M(q_{+}\bar{q}_{+})$\\
$T[(cq_{\pm})(\bar{c}\bar{q}_{\mp})]$& $\Rightarrow \hspace{-8pt}/$ (1st forbidden) &
$\psi(c\bar{c})+ M(q_{+}\bar{q}_{+})$ & 
${\epsilon(\mbox{\boldmath{P}})}^2\Gamma^{0}< \Gamma^{0}$ \\
$T[(cq_{-})(\bar{c}\bar{q}_{-})]$& $\Rightarrow\hspace{-12pt}//$ (2nd forbidden) &
$D(c\bar{q}_{+})+\bar{D}(\bar{c}q_{+})$ & ${\epsilon(\mbox{\boldmath{P}})}^4
\Gamma^{0} \ll \Gamma^{0}$\\
&&$\psi(c\bar{c})+ M(q_{+}\bar{q}_{+})$\\
\hline
\end{tabular}
\label{tab:TetraDecay}
\end{center}
\end{table}

The inspection of this qualitative decay properties, 
in addition to the other results in this section, led us to the 
assignments of the $X(3872)$ meson families shown in Fig.{\ref{fig:mass-spectra}}. 
Thus the $\rho_{3}$-rule seems to be able to explain the feature F2 of the new 
``exotic'' hadrons, mentioned in \S1. 
%And the above qualitative decay properties of tetra-quark systems 
%make \\

%%%%% N0 19  %%%%%%%%%%
\subsection{Near-Threshold Resonances observed at BES}

(Experimental Data)\\
~~~~Two Resonances{\cite{BES1}} with the following properties have been observed 
in $J/\psi \to \gamma +X$; 

$X[\omega\cdot \phi] \to \omega + \pi$: $M=1810 $MeV 
$\to$ close to $M_{th}(\omega +\pi)=1802$MeV, \\
~~~~~~$\Gamma=105$MeV, $(I,J^{PC})=(0,0^{++})$.\\

$X[p\cdot \bar{p}] \to p + \bar{p}$: $M=1859 $MeV 
$\to$ close to $M_{th}(p +\bar{p})=1876$MeV, \\
~~~~~~$\Gamma<28$MeV, $J^{PC}=0^{-+}$. \\

(Possible Structure in JSQM) \\
~~~~They seem to be identified as the 
four-quark $(qq\bar{q}\bar{q})$ and six-quark $(qqq\bar{q}\bar{q}\bar{q})$ systems, 
respectively 
with the structure shown in Fig.{\ref{fig:BESX}}
as follows: 
\vspace{-.5cm}
\begin{figure}[htbp]
\begin{center}
\includegraphics[width=13cm,height=6cm]{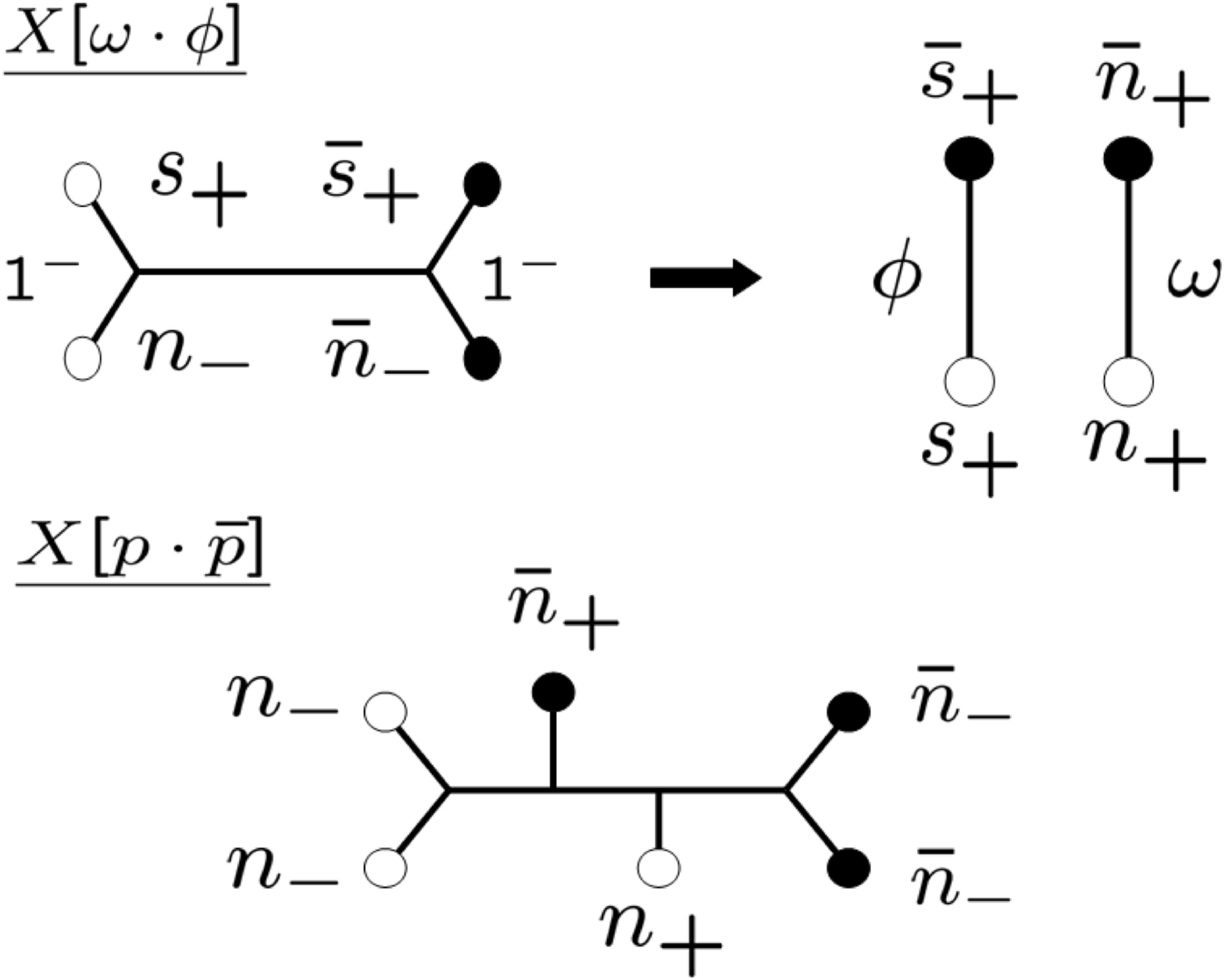}
\caption{Possible Structure of $X[\omega\cdot \phi]$ and $X[p\cdot \bar{p} ]$}
\label{fig:BESX}
\end{center}
\end{figure}

\begin{enumerate}
\item $X[\omega \cdot \phi]$\\
$J^{PC}$: $\underline{1}^{-}_{\chi} \otimes \underline{1}^{-}_{\bar{\chi}}=
\underline{0^{++}}, 1^{+-}, 2^{++}$\\
Mass: $M^{(0)}_{X}[\omega\cdot \phi]=M_{\phi}^{(0)}(s\bar{s})+M_{\omega}^{(0)}
(n\bar{n})$\\
%($\delta^{\chi} M_{X}$,$\delta^{J} M_{X} $ to be estimated)\\
Decey: $X[\omega\cdot \phi] \ \ {\Longrightarrow\hspace{-16pt}{//}}
 \ \  \phi (s_{+}\bar{s}_{+})+
\omega (n_{+}\bar{n}_{+}) $ ; doubly forbidden by the $\rho_{3}$-line rule. \\
Accordingly decay width to be $\Gamma =|\epsilon(P)|^4 \Gamma^{(0)} < \Gamma^{(0)}$.
\vspace{0.3cm}
\item $X[p \cdot \bar{p}]$\\
Mass: $M^{(0)}_{X}[p \cdot \bar{p}]=M_{p}^{(0)}(nnn)+M_{\bar{p}}^{(0)}
(\bar{n}\bar{n}\bar{n})$\\
%($\delta^{\chi} M_{X}$,$\delta^{J} M_{X} $ to be estimated)\\
Decey: $X[p \cdot \bar{p}] \ \ {\Longrightarrow\hspace{-9pt}{/}\hspace{-9.1pt}{/}
\hspace{-9.2pt}{/}\hspace{-9.3pt}{/}} \ \ \ \ \ 
N(n_{+} n_{+} n_{+})+
\bar{N} (\bar{n}_{+} \bar{n}_{+} \bar{n}_{+}) $ ; forbidden by the $\rho_{3}$-line rule 
in the fourth. \\
Accordingly $\Gamma =|\epsilon(P)|^8 \Gamma^{(0)} \ll \Gamma^{(0)}$.
\end{enumerate}
\vspace{0.3cm}
These tentative identification has been made, considering only the features F1 and F2 
of exotic hadrons (\S 1).
Detailed theoretical investigations are necessary to ascertain it. 

%%% NO 20 %%%%%%%%%%%

\section{Concluding Remarks}

We have reviewed the essentials of covariant framework in the $\widetilde{U}(12)$ 
classification scheme, where is predicted the existence of chiral states, new type 
of hadrons out of the conventional scheme. 
Hadron spectroscopy seems to be presently in serious difficulty, 
observing a new type of exotic hadrons successively. 
In going way out of this difficulty the notion of chiral states, 
closely connected to quark-confinement, 
is believed to play an important role. 
BES experiment is considered to be BEST place to seek for them. 
The serious correspondence to this situation be important and urgent!

%%%%%%%%%%%%%%%%%%%%%%%%%%%%%%%%%%%%%%%%%%%%%%%%%%%%%%%%%%%%%%%%%%%%%%%
%\end{flushleft}
%%%%%%%%%%%%%%%%%%%%%%%%%%%%%%%%%%%%%%%%%%%%%%%%%%%%%%%%%%%%%%%%%%%%%%%
\section*{Acknowledgements}
The author would like to express his sincere gratitude to Prof. Weiguo Li, 
Prof. X. Sheng, Dr. Wu Ning, and the other members of BES collaboration, who have 
supported this unique research-meeting. In this occasion he would newly thank 
Prof. K. Takamatsu and Prof. Z. P. Zheng for initiating our collaboration, in searching 
for the chiral states by use of BES data. He also thanks to Prof. H. Zehng for giving 
him a chance of seminar talk at Peking University. 
%%%%%%%%%%%%%%%%%%%%%%%%%%%%%%%%%%%%%%%%%%%%%%%%%%%%%%%%%%%%%%%%%%%%%%%
\appendix
\section{Present Picture of Hadrons and a History of Level-Classification}

A seminar talk, of which theme is closely related to that of this report, was given 
at the Peking University on February 24, 2006 by the present author. 
The title and contents 
were as follows: \\
 On the $\widetilde{U}(12)_{SF}$ classification Scheme and a ${U}(4)$ Symmetry for 
the Confined Light-Quarks; \ \ 
\S 0 Preliminaries,  \ \ 
\S 1 Present Status of Hadron Spectroscopy, \ \  
\S 2 Some Basic Consideration on Quark Confinement, \ \ 
\S 3 Essentials of Description of Composite Hadrons in $\widetilde{U}(12)$-Scheme, \ \  
\S 4 Concluding Remarks. \\
~~~~The contents of \S 1,\S 2 and \S 3, are almost the same as those in the present talk. 
Here I give a summary of the other parts not given here.
%the sections, \S0 preliminaries and \S4 Concluding Remarks.\\
\vspace{1em}
\bfl
(Proposal of $\widetilde{U}(12)_{SF}$-Level Classification Scheme)
\efl
~~~~The $\widetilde{U}(12)_{SF}$-classification scheme had been proposed{\cite{U12}} 
by us several years ago 
as a kinematical framework so as to correspond to the relativistic properties of 
composite hadrons. 
In this scheme we have seriously regarded the two general principles 
to be satisfied by any physics theory: \\
%In this talk I shall review the $\widetilde{U}(12)$-level classification scheme 
%of hadrons, which had been proposed by ourselves several years ago. 
%In this scheme we have regarded seriously the two general principles: \\

P-E. \ \ Lorentz Covariance \ \ \ in describing the space-time structure of 
composite hadrons.

P-H. \ \ Observable Notion \ \ \ in connection to the confinement of constituent quarks. 
\\

\bfl
(Non-Relativistic Picture of Hadrons and Facing Problem)
\efl
~~~~The hadron is conventionally regarded as the color-singlet bound-states of quarks 
and anti-quarks, where the constituent quarks seem to obey the Non-Relativistic 
quantum mechanics and have approximately the $SU(6)_{SF} \otimes O(3)_{L}$ symmetry. 
On the contrary, the composite hadrons are surely relativistic entities, since 
the pion, for example, has the properties of Nambu-Goldstone boson in the case of 
spontaneously broken-chiral symmetry. Accordingly we are facing the problem of extending 
the above NR picture relativistically. \\

\bfl
(Difficulty of Relativistic Extension)\\
\efl
~~~~One the most natural covariant-extension of $SU(6)_{SF} \otimes O(3)_{L}$ scheme 
is thought to be the $\widetilde{U}(12)_{SF} \otimes O(3,1)_{\rm Lorentz}$, treating 
separately the external (center of mass) and the internal (relative) coordinates of 
composite hadrons. However, this way of extension seems to be almost closed: \\
~~~~On the one hand, the original theory{\cite{R26}} of $\widetilde{U}(12)_{SF}$ symmetry, 
which was the first attempt of the extension of $SU(6)_{SF}$, had been shown to have the 
serious difficulty as follows. \\
~~~~First the physical state condition (restricting within the boosted-Pauli spinors) 
reduces to the violation of unitarity{\cite{R27}}, 
and secondly the ``no-go-theorem
{\cite{R22}}'' states that a relativistic extension of the $SU(6)_{SF}$ symmetry is 
impossible. \\
~~~~On the other hand, in the framework of bilocal-field theory by Yukawa 
(,which is regarded as the 
first attempt considering covariantly the spatial-extension of 
composite system, ) it had been shown 
that the framework leads to the violation of general principle such as causality, 
unitarity and even Lorentz-covariance. This implies also 
another ``no-go-theorem'' against the above way of extension. \\
~~~~Here we remind ourselves (consciously) that there exists, presently, no consistent 
non-local (to represent the internal spatial-extension), quantum field theory. 
As a matter of fact that various attempts for relativistic generalization 
of the above hadron 
picture ever-appeared have anyhow some unsatisfactory points as follows : \\
~~~~The Bethe-Salpeter equation for relativistic bound-state problem starts from the 
asymtotically free quark states as the S-matrix bases, 
being against the principle P-H. \\
~~~~The heavy quark 
effective theory is still a static theory, being against the princilple P-E, 
because the heavy-quark stays at rest. \\
~~~~The method of effective Lagrangian is, 
able to treat the chiral symmetry, but the framework is a local field 
theory, unable to treat the spatial extension of hadrons.\\

\bfl
(Exciton Picture of Quarks and Static $U(12)_{SF}$ Symmetry in the
$\widetilde{U}(12)$-scheme)
\efl
~~~~In the new $\widetilde{U}(12)_{SF}$-classification scheme 
the constituent quark is regarded 
as, not an elementary entity of bound state hadrons but the exciton{\cite{Ur}}, 
which is a 
mathematical entity simulating the center of mass motion of confined quarks inside 
relevant hadrons. \\
~~~~The $U(12)_{SF}(\supset U(4)_{S} \otimes SU(3)_{F})$ symmetry is embedded
{\footnote{As an origin of this thought I should like to refer to the $SU(6)_{SF}$
symmetry as a ``rest condition''{\cite{R14}}.} in the 
covariant $\widetilde{U}(12)_{SF} \otimes O(3,1)_{\rm Lorentz}$ space in the hadron 
rest-frame. Accordingly our new scheme is out of 
the above mentioned ``no-go-theorem''. \\
~~~~Furthermore, it is also free from the trouble related to unitarity, since 
there is now no need for the physical-state condition. 
The ``unphysical'' states 
(at that time), described with chiral(urciton)-spinors, are now considered to be 
an origin of 
promising candidates of new ``exotic hadrons''out of the conventional NR-classification 
scheme.\\
~~~~In the Table {\ref{tab:nen}} 
a brief history of the hadron symmetry and classification, 
in relation to the $\widetilde{U}(12)$-classification scheme, is given.\\
%%%%%%%%%%%%%%%%%
\let\tabularsize\scriptsize
\begin{table}[t]
\caption{A Histrical Road led to $\widetilde{U}(12)$-Classification Scheme}
\begin{tabular}{lll}
\hline
\hline
& & \\
1950&H. Yukawa  & Bi-local Field Theory; To avoid divergence.\\
    &           & Wave Function $\Phi(x_{\mu}, r_{\mu})$ violates, in higher orders, \\
    &           & covariance, causality and unitarity(C. Hayashi).\\
1953&H. Yukawa  & Covar. Oscillator WF; Freedom on $r_{\mu}$ gives mass spectra of mesons.\\
    &           & $\to$ Oscillator Model $\sim$1970 (Sogami, Feynman, Namiki, COQM).\\
1956&S. Sakata  &Composite Model; Origin of flavor $[q_{a}:a=(1,2,3)]$ \\
    &           &$\to$ $SU(3)_{F}$ symmetry $\sim$ 1959 (Ikeda-Ogawa-Ohnuki, Yamaguchi).\\
1964&M. Gell-Mann  & Quark Model; \\
    &G. Zweig      & $\to$ $SU(6)_{SF}$ symmetry
$\sim$ 1964 (F. Gursey and L.A.Radicati, B. Sakita).\\
    &              &; Addition of $SU(2)_{\sigma}$-spin symmetry$[q_{a,i}:i=(1,2)]$.\\
1965&A. Salam et al. & ``Original'' $\widetilde{U}(12)$-symmetry; \\
    &B. Sakita-K.C.Wali & Attempt for relativ. ext. of $SU(6)_{SF}$ symmetry
$[q_{a,\alpha}:\alpha=(1 \sim 4)]$.\\
&& $\widetilde{U}(12)_{SF} \supset SU(3)_{F} \otimes \widetilde{U}(4)_{\rm Dirac Spinor}$ with physical 
state condition.\\
&& Physical states be Boosted-Pauli spinors $\to$ ``Violation of unitarity.''\\
1967&S. Coleman& ``No-Go Theorem'';\\
 &-J. Mandula& Lorentz covariant symm. including $SU(6)_{SF} \supset SU(3)_{F}
    \otimes SU(2)_{\sigma}$,\\
    && not existing.  \\
1968&S. Ishida-P. Roman & ``$SU(6)_{SF}$ Symm.'' as ``Rest Condition''.\\
%   &   &\\
1970&S. Ishida, M. Oda, &Urciton Scheme; urciton, simulating confined quarks, \\
& K. Yamada& WF $\Phi_{A}{}^{B}(X_{\mu}, r_{\mu}) \ \ A=(a, \alpha)$ ; Tensors in \\
&& $\widetilde{U}(12)_{SF} \otimes O(3,1)$ space with $SU(6)_{SF} \otimes O(3)_{L}$ symm.
``at Rest''.\\
$\sim$&& The kinematical framework (still keeping P.S. condition and suffering \\
&& from the ``violation of unitarity.''), COQM, had been applied  \\
&& to Born term of hadron-reactions. \\
2000& S.Ishida-M.Ishida& New $\widetilde{U}(12)$ level classification 
scheme~; \\
 &-T.Maeda&Static $U(12)_{SF}\otimes O(3)_{L}$ symm. embedded in $\widetilde{U}(12)_{SF} \otimes O(3,1)$ space.  \\
2002&S.Ishida-M.Ishida& Free from ``No-Go Theorem'' and ``Violation of Unitarity''
\\
& & \\
\hline
\end{tabular}
\label{tab:nen}
\end{table}
%%%%%%%%%%%%%%%%%
\normalsize
\bfl
(Implication of Possible Existence of Chiral States)
\efl
~~~~In the $\widetilde{U}(12)$-classification scheme the existence of chiral states, 
which are out of the conventional NR classification scheme, is predicted. \\
~~~~Their confirmation would imply the {\it realization in nature} of 
{\it the lowest and the higher-number-index solutions} 
of the original Dirac Equation {\it as representing the leptons and the 
hadrons}, 
respectively. The equation had been proposed long ago, as is well known, in order to 
derive the Lorentz-covariant (first-quantized) wave equation without negative-energy 
solutions.
%%%%%%%%%%%%%%
%(Hadron picture and facing problem of hadron physics)\\
%
%The hadron is presently considered to be bound states of quarks 
%with the quantum numbers, flavor, spin, and color. The constituent quark, as is 
%suggested from the simple quark model, seems to be obey approximately 
%the non-relativistic quantum mechanics, and is confined inside hadrons as 
%color-singlet states. On the other hand, the $\pi$-meson, a typical composite 
%hadron behaves phenomenologically like the Nambu-Goldstone boson 
%in the case of spontaneously broken-chiral symmetry. But non-relativistic theory is 
%out of notion of the chiral symmetry since its generator $\gamma_{5}=\gamma_{1}
%\gamma_{2}\gamma_{3}\gamma_{4}$, which being Lorentz-covariant, is missing there. 
%Accordingly the above picture produced a serious problem in hadron physisc, 
%because there exists, presently, no consistent non-local (representing 
%the composite structure) and relativistically covarinat quantum field theory.\\


\newpage

\begin{thebibliography}{99}
%%%%%%%%%%%%%%%%%%%%%%%%%%%%%%%%%%%%%%%%%%%%%%%%%%%%%%%%%%%%%
% Some macros are available for the bibliography:
%  o for general use
%    \JL : general journals                 \andvol : Vol (Year) Page
%  o for individual journal 
%    \AJ   : Astrophys. J.           \NC         : Nuovo Cim.
%    \ANN  : Ann. of Phys.           \NPA, \NPB  : Nucl. Phys. [A,B]
%    \CMP  : Commun. Math. Phys.     \PLA, \PLB  : Phys. Lett. [A,B]
%    \IJMP : Int. J. Mod. Phys.      \PRA - \PRE : Phys. Rev. [A-E]     
%    \JHEP : J. High Energy Phys.    \PRL        : Phys. Rev. Lett.
%    \JMP  : J. Math. Phys.          \PRP        : Phys. Rep.
%    \JP   : J. of Phys.             \PTP        : Prog. Theor. Phys.     
%    \JPSJ : J. Phys. Soc. Jpn.      \PTPS       : Prog. Theor. Phys. Suppl.
% Usage:
%  \PRD{45,1990,345}          ==> Phys.~Rev.\ \textbf{D45} (1990), 345
%  \JL{Nature,418,2002,123}   ==> Nature \textbf{418} (2002), 123
%  \andvol{B123,1995,1020}    ==> \textbf{B123} (1995), 1020
%%%%%%%%%%%%%%%%%%%%%%%%%%%%%%%%%%%%%%%%%%%%%%%%%%%%%%%%%%%%%
\bibitem{Swanson} E. S. Swanson, hep-ph/0601110.
\bibitem{U12} S. Ishida, M. Ishida, \PLB{539,2002,249}. \\
S. Ishida, M. Ishida, and T. Maeda, \PTP{104,2000,785}
\bibitem{Ds} S. Ishida et al., hep-ph/0408136.
\bibitem{Ur} S. Ishida, \PTP{46,1971,1570;1905}.
\bibitem{Watanabe} S. Watanabe, \PR{106,1957,1306}.
\bibitem{Sakurai} J. J. Sakurai, \NC{7,1958,649}. \ 
J. Tiomno, \NC{1,1955,226}.
\bibitem{Yukawa} H. Yukawa, \PR{91,1953,415}. See also the related works{\cite{PTP}}.
\bibitem{PTP} H. Takabayashi, \NC{33,1964,668},\\
I. Sogami,\PTP{41,1969,1352},\\
K. Fujimura, T. Kobayashi and M. Namiki, \PTP{43,1970,73},\\
Y. S. Kim and M. E. Noz, \PRD{8,1973,3251}.
\bibitem{protovino} S. Ishida and M. Ishida, in proc. of HADRON'01(Protovino), 
hep-ph/110359.
\bibitem{Imachi} M. Imachi et. al, \PTP{55,1976,551}.\\
F. Toyoda, M. Uehara, \PTP{58,1977,1456}.\\
S. Ishida et. al, \PTP{68, 1982, 883}.
\bibitem{ourX} M. Ishida et al., hep-ph/0509212, S. Ishida, hep-ph/0511014.
\bibitem{BES1} S.Jin, Talk at HADRON'05, Rio de Janeiro.\\
 J.Z. Bai et al., BES collaboration, \PRL{91,2003,022001}.
\bibitem{R26} A. Salam, R. Delbourgo and J. Strathdee, Proc. Roy. Soc.(London)
\textbf{A284} (1965), 146.
\bibitem{R27} S. Weinberg, \PR{139,1965,146}.
\bibitem{R22} S. Coleman and J. Mandula, \PR{177,1969,2371}.
\bibitem{R14} S. Ishida, P. Roman, \PR{172,1968, 610}
\bibitem{R15} P. A. M. Dirac, Proc. Roy. Soc.(London) \textbf{A117} (1928), 620.
\bibitem{R} S. Ishida, M. Ishida and M. Oda, \PTP{93,1995,939}. 
\end{thebibliography}
\end{document}